\documentclass[pss]{wiley2sp} 
\usepackage{amsmath}
\usepackage{siunitx}
\usepackage{hyperref}

\DeclareSIUnit\micron{\micro\metre}
\DeclareSIUnit\nb{\nano\barn}
\DeclareSIUnit\pb{\pico\barn}
\DeclareSIUnit\fb{\femto\barn}
\DeclareSIUnit\particle{p}

\begin{document}

\title{Severe signal loss in diamond beam loss monitors in high particle rate environments by charge trapping in radiation induced defects}

\titlerunning{Diamond polarization}

\author{%
  Florian Kassel\textsuperscript{\Ast,\textsf{\bfseries 1}},
  Moritz Guthoff\textsuperscript{\textsf{\bfseries 2}},
  Anne Dabrowski\textsuperscript{\textsf{\bfseries 2}},
  Wim de Boer\textsuperscript{\textsf{\bfseries 1}}}

\authorrunning{Florian Kassel et al.}

\mail{e-mail
  \textsf{florian.kassel@cern.ch}, Phone: +41 75 411 7486}

\institute{%
  \textsuperscript{1}\,Institute for Experimental Nuclear Physics (IEKP), KIT, Karlsruhe, Germany\\
  \textsuperscript{2}\,CERN, Meyrin, Switzerland}

\received{21 March 2016, revised 30 May 2016, accepted 9 June 2016} 
\published{18 July 2016 in Physica Status Solidi (a), 1-9 (2016) / \href{http://onlinelibrary.wiley.com/doi/10.1002/pssa.201600185/full}{DOI 10.1002/pssa.201600185}} 


\abstract{%
%
%
%
\abstcol{%
  The Beam Condition Monitoring Leakage (BCML) system is a beam monitoring device in the CMS experiment at the LHC. As detectors 32 poly-crystalline (pCVD) diamond sensors are positioned in rings around the beam pipe. Here high particle rates occur from the colli\-ding beams scattering particles outside the beam pipe. These particles cause defects, which act as traps for the ionization, thus reducing the charge collection efficiency (CCE). However, the loss in CCE was much more severe than expected from low rate laboratory measurements and simulations, especially in single-crystalline (sCVD) diamonds, which have a low initial concentration of defects. After an integrated luminosity of a few \si{\per\fb} corresponding to a few weeks of LHC o\-per\-ation, the CCE of the sCVD diamonds dropped by a factor of five or more and quickly approached the poor CCE of pCVD diamonds. The reason why in real experiments the CCE is much worse than in laboratory experiments is related to the ionization rate.
  }{%
  At high particle rates the trapping rate of the ionization is so high compared with the detrapping rate, that space charge builds up. This space charge reduces locally the internal electric field, which in turn increases the trapping rate and recombination and hence reduces the CCE in a strongly non-linear way. A diamond irradiation campaign was started to investigate the rate dependent electrical field deformation with respect to the radiation damage. Besides the electrical field measurements via the Transient Current Technique (TCT), the CCE was measured. The experimental results were used to create an effective deep trap model that takes the radiation damage into account. U\-sing this trap model the rate dependent electrical field deformation and the CCE were simulated with the software \mbox{SILVACO TCAD}. The simulation, tuned to rate dependent measurements from a strong radioactive source, was able to predict the non-linear decrease of the CCE in the harsh environment of the LHC, where the particle rate was a factor 30 higher.}}

%
%
%

\maketitle   

\begin{figure}[tbh]%
\includegraphics*[width=\linewidth]{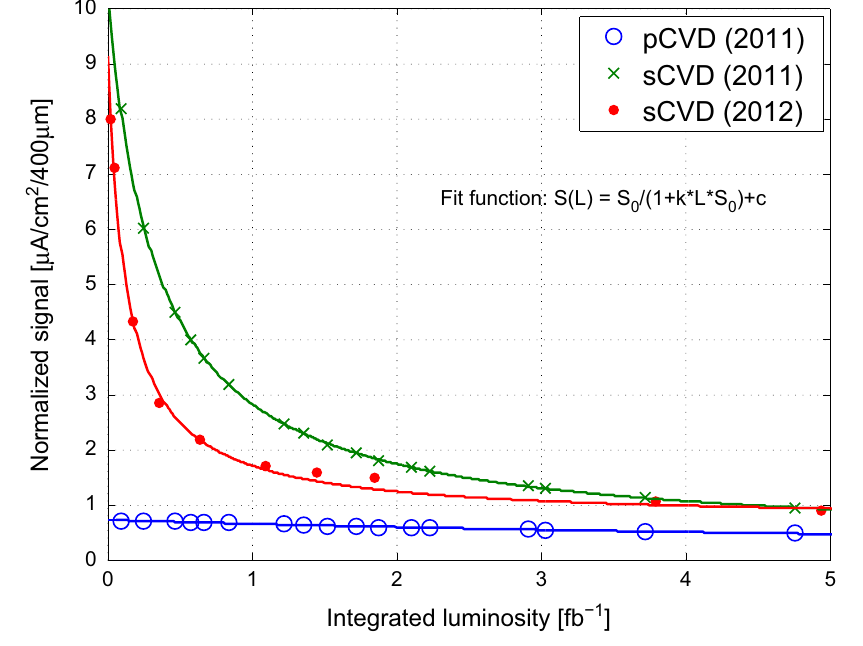}
\caption{%
  Signal loss of two sCVD diamond sensors (green and red) compared to the signal of a pCVD diamond (blue) \cite{Guthoff2013168}.}
\label{BCML_signalloss}
\end{figure}

\begin{figure}[tbh]%
\includegraphics*[width=\linewidth]{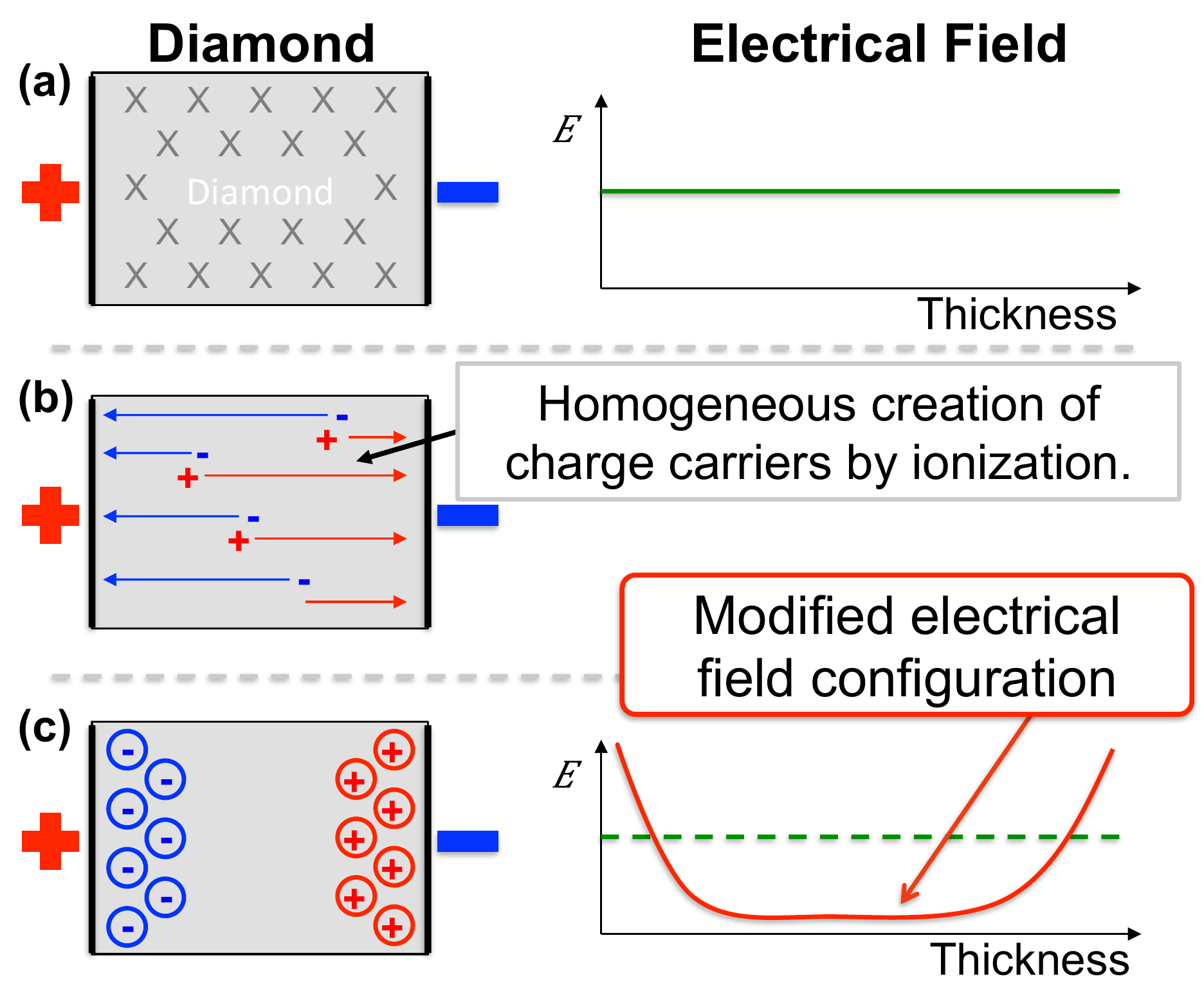}
\caption{%
  Diamond polarization: Radiation damage leads to a homogeneous distribution of lattice defects in the diamond bulk (a). The external electrical field leads to an asymmetrical distribution of the charge carrier density, since the flux and hence the trapping probability of positive (negative) charge carriers is proportional to the thickness of the collecting region, as shown in (b). This results in an asymmetrical build-up of space charge (c) and hence in a locally reduced electrical field that causes an increase in charge carrier recombination and therefore a reduced charge collection efficiency.}
\label{Polarization_sketch}
\end{figure}

\section{Introduction}
The CMS Beam Condition Monitor Leakage (BCML) system at LHC is a beam monitoring device based on 32 poly-crystalline (pCVD) diamond sensors which detects the ionization from the beam losses leaking outside the beam pipe, e.g. by scattering on the residual gas or beam collimators. Although diamond sensors were expected to be radiation hard, the charge collection efficiency (CCE) dropped much faster \cite{Guthoff2013168,Guthoff2015,Guthoff2014} than expected from low particle rate laboratory measurements \cite{RD42} and simulations \cite{deBoer2007,Guthoff2013}, especially in single-crystalline (sCVD) diamonds, which have a low initial concentration of defects. After an integrated luminosity of a few \si{\per\fb} corresponding to a few weeks of LHC operation, the CCE of the sCVD diamonds dropped by a factor of five or more and quickly approached the poor CCE of pCVD diamonds, see Fig.~\ref{BCML_signalloss}. This discrepancy in CCE between the real application in a particle detector and laboratory experiments can be explained by the rate dependent polarization \cite{Rebai2016,Valentin2015} of the \mbox{diamond} \mbox{detector,} as was deduced from detailed laboratory measurements and simulations.

\paragraph{Rate dependent diamond polarization}
Radiation damage introduces a homogeneous distribution of defects in the diamond bulk, which can act as traps for charge carriers created by ionization. With an external electrical field applied the electrons are drifting to the anode and the holes are drifting to the cathode. The charge trapping is linear proportional to the charge carrier density and hence leads to an increased hole trapping at the cathode and an increased electron trapping at the anode. This creation of space charge predominantly towards the edges leads to a mo\-di\-fied electrical field distribution inside the diamond bulk with a local minimum in the central region (see Fig.~\ref{Polarization_sketch}). In this low field region the recombination rate is increased and hence the charge collection efficiency reduced. At high particle rates the trapping rate is higher than the detrapping rate, so space charge builds up even stronger and reducing the CCE more.

\begin{table}
\centering
  \caption{Overview of the irradiated diamond samples. The initial CCE was measured at an electrical field of $E = $\SI{1800}{\volt\per\square\centi\meter}. The sensors were either irradiated with \SI{23}{\mega\electronvolt} protons (p) or with neutrons (n) of an energy distribution up to \SI{10}{\mega\electronvolt}  for \cite{NeutronFacility}. The damage factors are 2.8 for protons (NIEL \cite{Guthoff2013}) and 1.2 for neutrons (NIEL \cite{Cindro2015}) to \SI{1}{\mega\electronvolt} neutron equivalent.}
  \begin{tabular}[htbp]{@{}ccccc@{}}
    \hline
    Name & Thickness & CCE$_{\rm{ini}}$ & Irradiation & Tot. fluence\\
    & (\si{\micron}) &  (\si{\percent}) & type & ($\times$ \SI{e12}{\per\square\centi\metre}) \\
    \hline
    Sample \#1 & 549 & 98 & p & 3 \\
    Sample \#2 & 549 & 95 & p & 2 \\
    Sample \#3 & 538 & 98 & n & 1\\
    Sample \#4 & 542 & 92 & n & 1\\
    \hline
  \end{tabular}
  \label{Diamond_parameters}
\end{table}

\section{Experimental study of the diamond polarization}
Four new single crystalline diamonds of highest quality 'electronic grade' corresponding to $\left[N\right] < 5\,\rm{ppB}$ and $\left[B\right] < 1\,\rm{ppB}$ produced by Element6 \cite{Element6} were used to investigate the diamond polarization with respect to the irradiation damage. On the $5\times 5\,\rm{mm}$ diamond surfaces ohmic titanium/tungsten electrodes were sputtered with a thickness of $100\,\rm{nm}$ after $1\,\rm{\mu m}$ of the surfaces was removed by a chlorine chemistry reactive ion etching. The metal electrodes with a surface of $4\times 4\,\rm{mm}$ were annealed in an $N_2$ environment for $4\,\rm{min}$ at $400\,\rm{^{\circ}C}$.

The diamond samples were irradiated stepwise with $23\,\rm{MeV}$ protons (sample $\#1$ and $\#2$) and with neutron particles with an energy distribution up to $10\,\rm{MeV}$ \cite{NeutronFacility} (sample $\#3$ and $\#4$) to a maximum fluence of $3\times 10^{12}\,\rm{cm^{-2}}$. An overview of the irradiated samples is given in Table \ref{Diamond_parameters}.

Diamond polarization is modifying the internal electrical field, which can be measured using the Transient-Current-Technique (TCT) \cite{Isberg2002,Pernegger2005}. The electrical field modification is affecting also the sensor efficiency, therefore CCE measurements were done as well. In the subsections \ref{TCT_setup_chapter} and \ref{CCE_chapter} both measurement techniques will be introduced. The basic concept measuring the build-up of polarization is discussed in section \ref{Measurement_procedure_chapter}.

\begin{figure}[tbh]%
\includegraphics*[width=\linewidth]{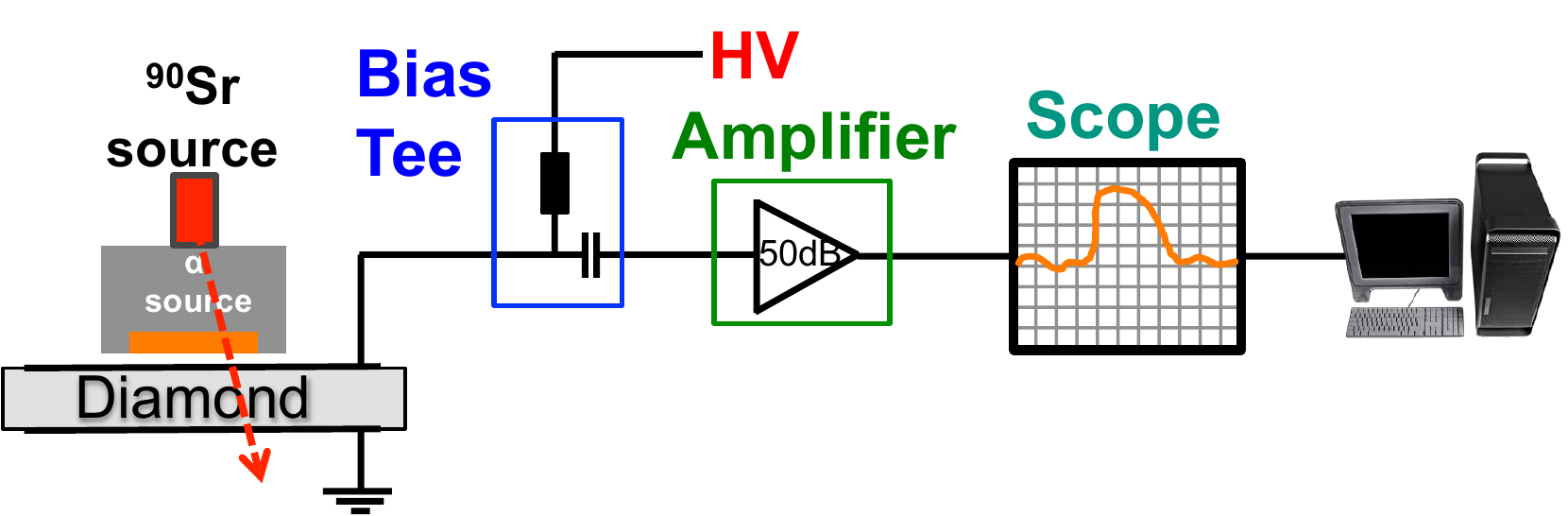}
\caption{%
  The TCT setup consists of a bias tee, a signal amplifier and an oscilloscope. The TCT signal created by $\alpha$ particles is decoupled from the high voltage circuit into the signal circuit via the bias tee. The high frequency TCT signal is amplified and read out with a digital storage scope with a bandwidth of $1\,\rm{GHz}$.}
\label{TCT_setup}
\end{figure}

\subsection{Transient-Current-Technique (TCT)}\label{TCT_setup_chapter}
The electrical field can be measured indirectly using TCT: electron-hole pairs are induced on one side of the diamond by an $\alpha$ particle. The external applied electrical field causes the charge carries to drift in opposite direction through the diamond bulk. The small penetration depth of the $\alpha$ particle $\leq 15\,\rm{\mu m}$ (section.~\ref{simulation_desc_chapter}) causes however one type of charge carrier to reach immediately the electrode. The other type of charge carrier is drifting through the entire diamond bulk. Based on Ramo's theorem \cite{Ramo1939} drifting charge is in\-du\-cing a current in the electrical circuit. This current as function of time is proportional to the electric field as function of distance. Hence for a constant electrical field a square TCT signal should be obtained. Creating the electron-hole pairs on the cathode or anode side allows measuring the electron or the hole drift, respectively.

The TCT setup shown in Fig.~\ref{TCT_setup} consists of a Picosecond 5531 bias tee, a Particulars wide band current amplifier ($53\,\rm{dB}$, $0.01-2\,\rm{GHz}$) \cite{Particulars_Amp}, a Tektronix TDS 5104B oscilloscope ($5\,$GSPS, $1\,$GHz) and a Keithley 2410 as bias voltage source. The limiting component in terms of bandwidth is the oscilloscope with $1\,$GHz. A $3.56\,\rm{kBq}$ $^{241}Am$ source is used to create $\alpha$ particles. In addition $\beta$ particles from a $32.2\,\rm{MBq}$ $^{90}Sr$ source are used to create a constant ionization rate, creating electron-hole pairs homogeneously in the entire diamond bulk.

The $\alpha$ particle is stopped inside the material, therefore an external trigger can't be used. The read-out by the digital storage scope is triggered internally with a threshold of $25\,\rm{mV}$. The TCT pulses from the scope are read out with a frequency of $2.5\,\rm{Hz}$. The post processing of the data consists of baseline subtraction, realignment and averaging.

\subsection{CCE measurement}\label{CCE_chapter}
The signal created by a mi\-nimum ionizing particle (MIP) from a $^{90}Sr$ source pas\-sing through the entire diamond bulk is used to calculate the diamond CCE. A MIP particle creates 36.7 electron-hole pairs per $\rm{\mu m}$ of diamond thickness. Using the diamond thickness the total induced charge $Q_{\rm{induced}}$ can be calculated. The measured charge $Q_{\rm{meas}}$ of the MIP particle signal is compared to the induced charge $Q_{\rm{induced}}$ to calculate the detector efficiency, following Eq.~\ref{CCE_calc_eq}:
\begin{equation}
\label{CCE_calc_eq}
\rm{CCE} = \frac{Q_{\rm{meas}}}{Q_{\rm{induced}}}.
\end{equation}
The MIP particles are created by a $33.4\,\rm{MBq}$ $^{90}Sr$ source, which creates simultaneously a constant ionization rate. A detailed description of the CCE setup can be found in \cite{Grah2009}.

\subsection{Measurement procedure}\label{Measurement_procedure_chapter}
In order to measure the build-up of polarization in the diamond the following procedure is used for the TCT and CCE measurements:
\begin{enumerate}
	\item The diamond is exposed during the entire measurement to a constant ionization rate by a $^{90}Sr$ source creating electron-hole pairs in the entire diamond bulk filling up the traps. In the steady-state the trapping and detrapping are in equilibrium. According to the simulations discussed below about 55\,\% of the effective deep traps are filled.
	\item In order to remove any residual field and set the diamond into a unpolarized state, the sensor is exposed to the $^{90}Sr$ source for a duration of 20\,minutes without bias voltage applied. A homogeneous trap filling in the diamond bulk, and hence an unpolarized diamond state, is reached.
	\item The bias voltage is ramped up fast ($t_{\rm{ramp}}\leq 10\,\rm{s}$) and the measurement is started immediately ($t=0\,\rm{s}$).
	\item With bias voltage applied the diamond starts to polarize. The measurement is performed over an extended period of time ($t>3000\,\rm{s}$) until the diamond is fully polarized and the measurement results are stable.
\end{enumerate}

\begin{figure*}%
\subfloat[Hole drift: undamaged]{%
\includegraphics*[width=.3\textwidth]{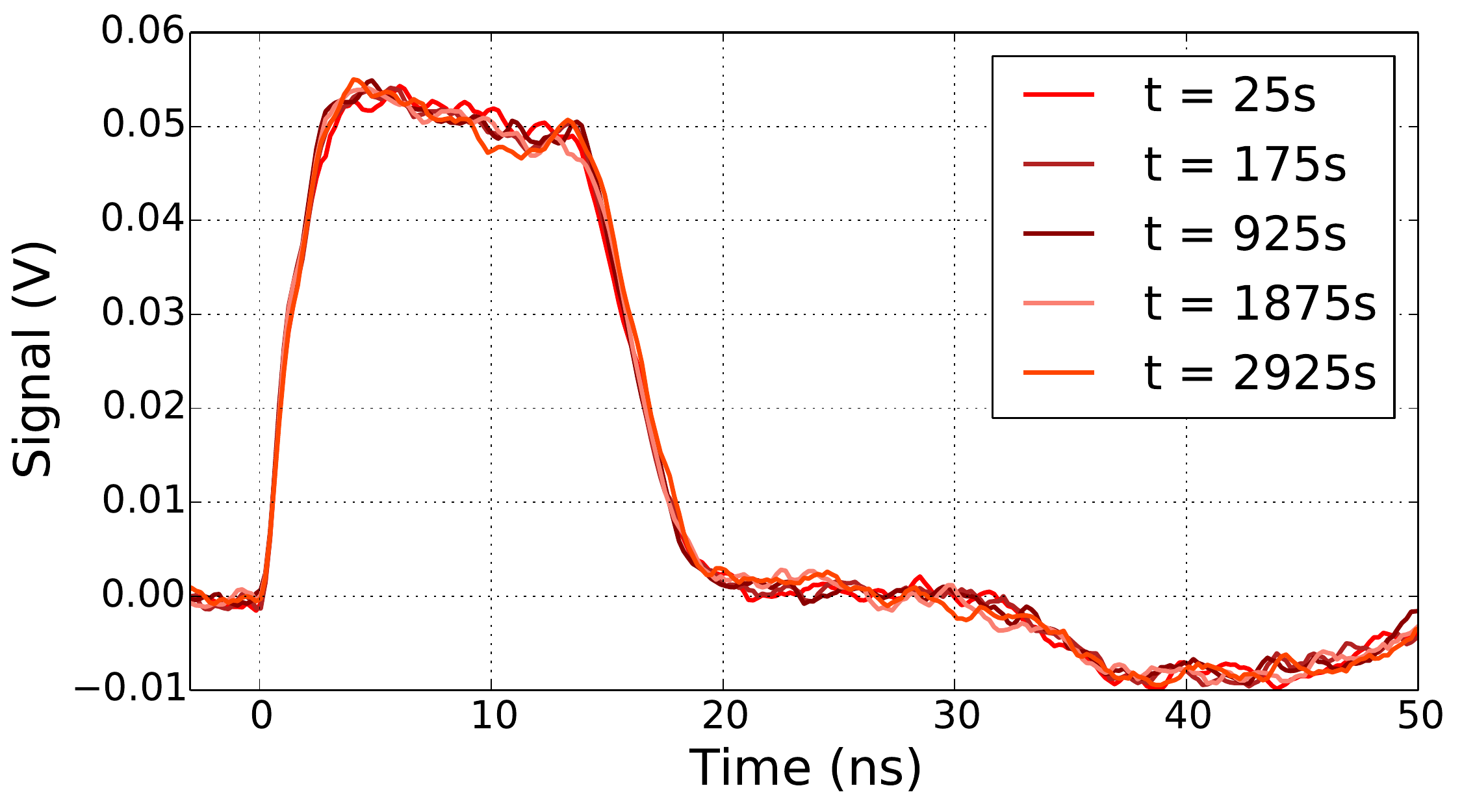}\label{TCT_f0_hole_1}}\hfill
\subfloat[Hole drift: $f_1=1\times 10^{12}\,\rm{cm}^{-2}$]{%
\includegraphics*[width=.3\textwidth]{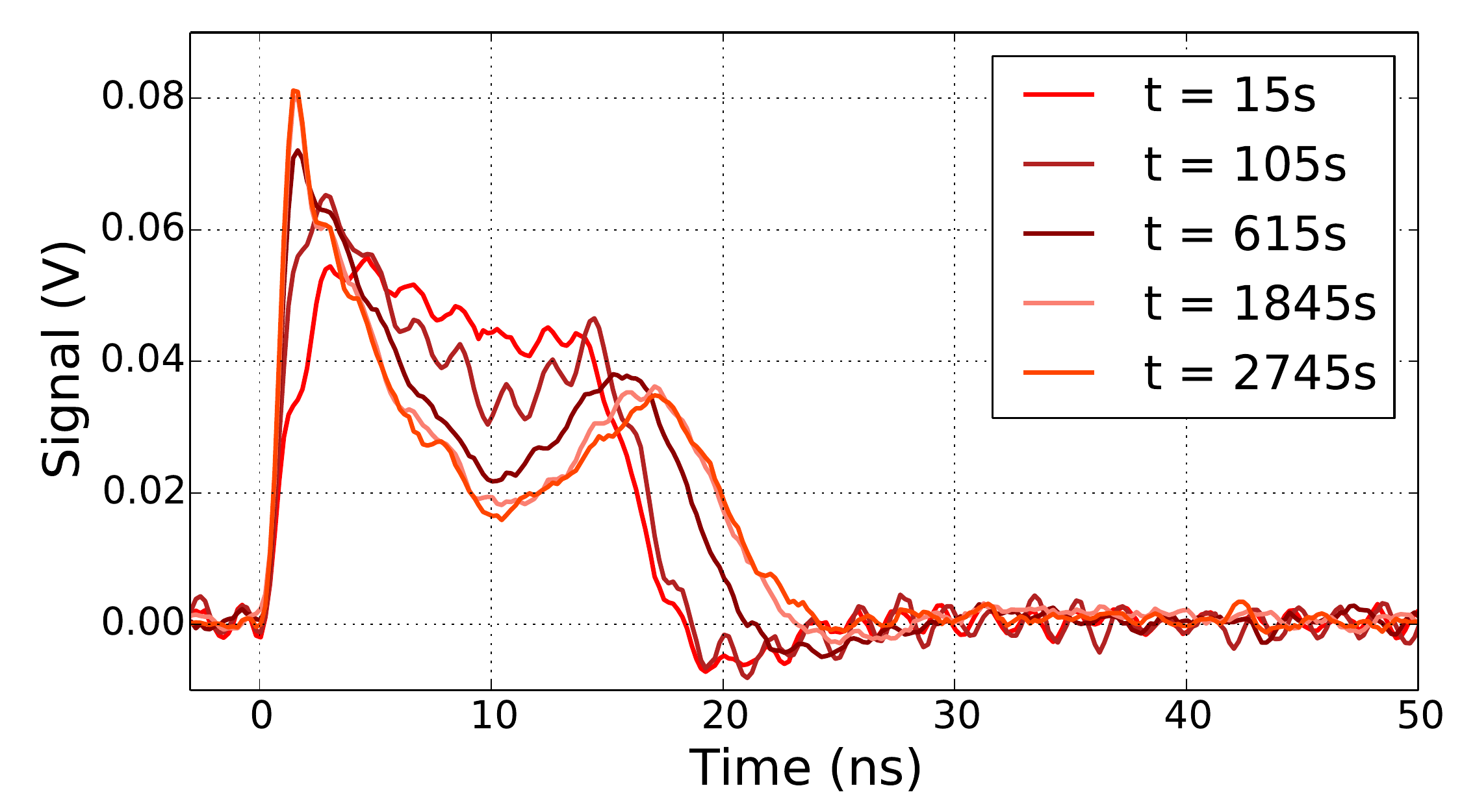}\label{TCT_f0_hole_2}}\hfill
\subfloat[Hole drift: $f_2=3\times 10^{12}\,\rm{cm}^{-2}$]{%
\includegraphics*[width=.3\textwidth]{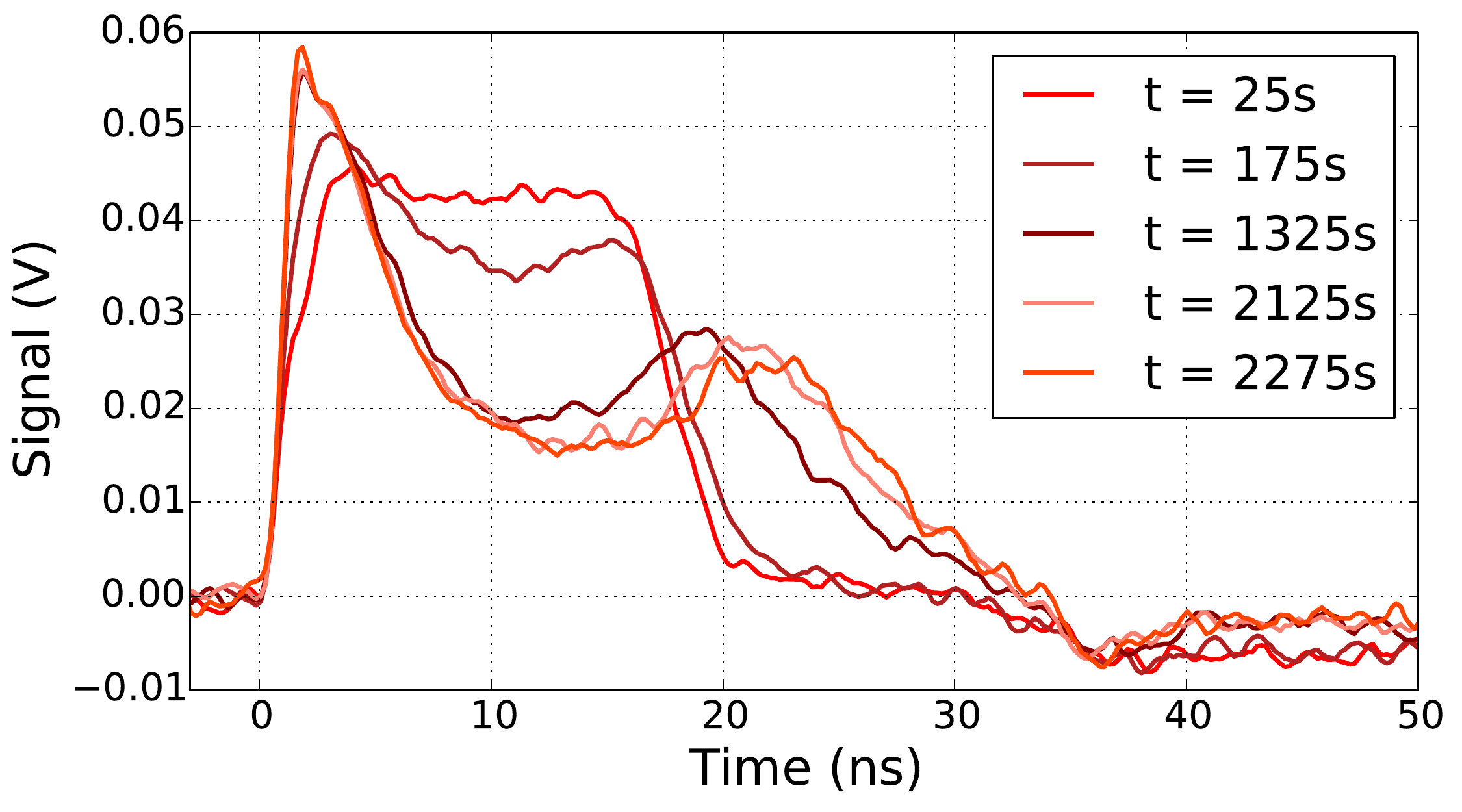}\label{TCT_f0_hole_3}}\hfill
\subfloat[Electron drift: undamaged]{%
\includegraphics*[width=.3\textwidth]{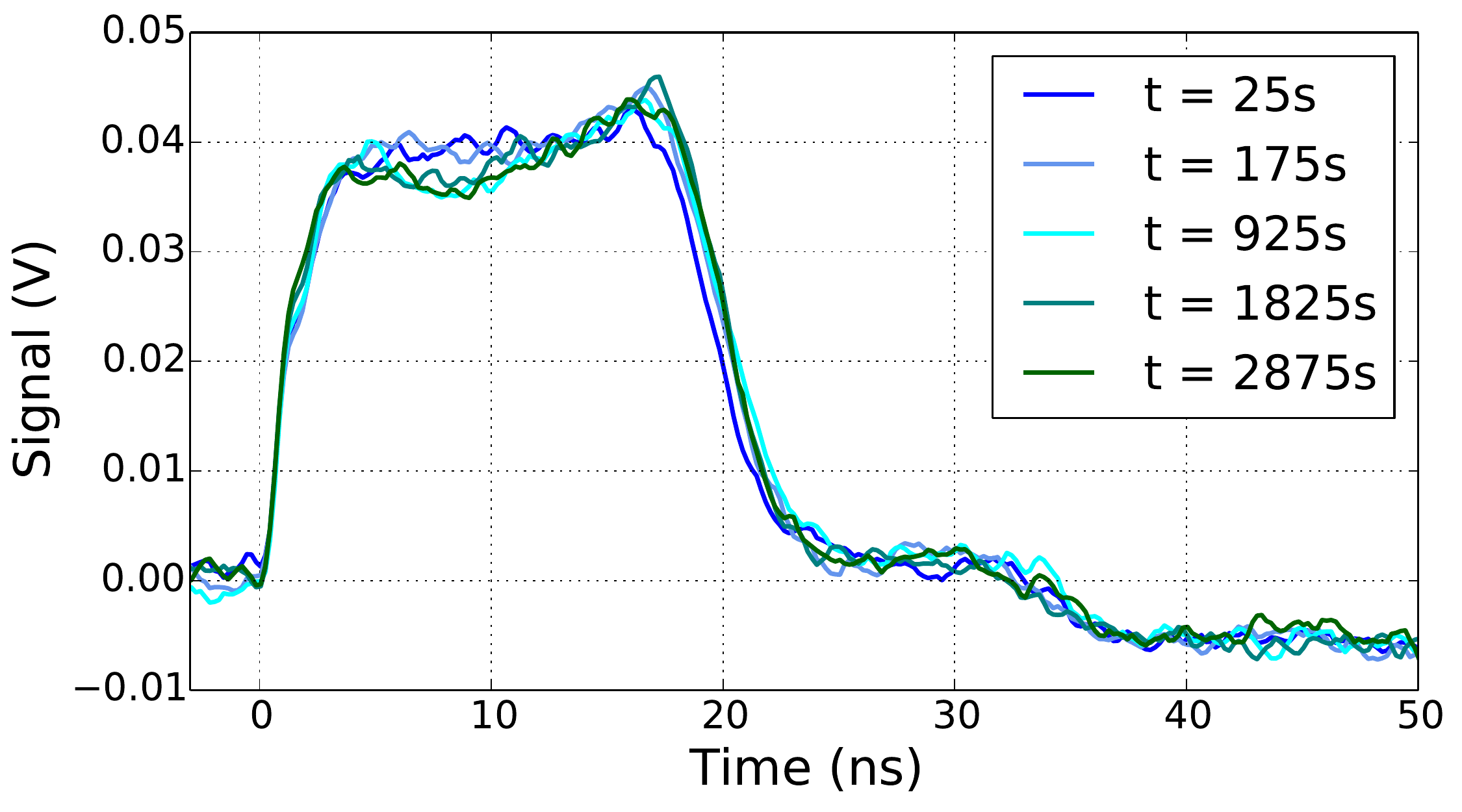}\label{TCT_f0_ele_1}}\hfill
\subfloat[Electron drift: $f_1=1\times 10^{12}\,\rm{cm}^{-2}$]{%
\includegraphics*[width=.3\textwidth]{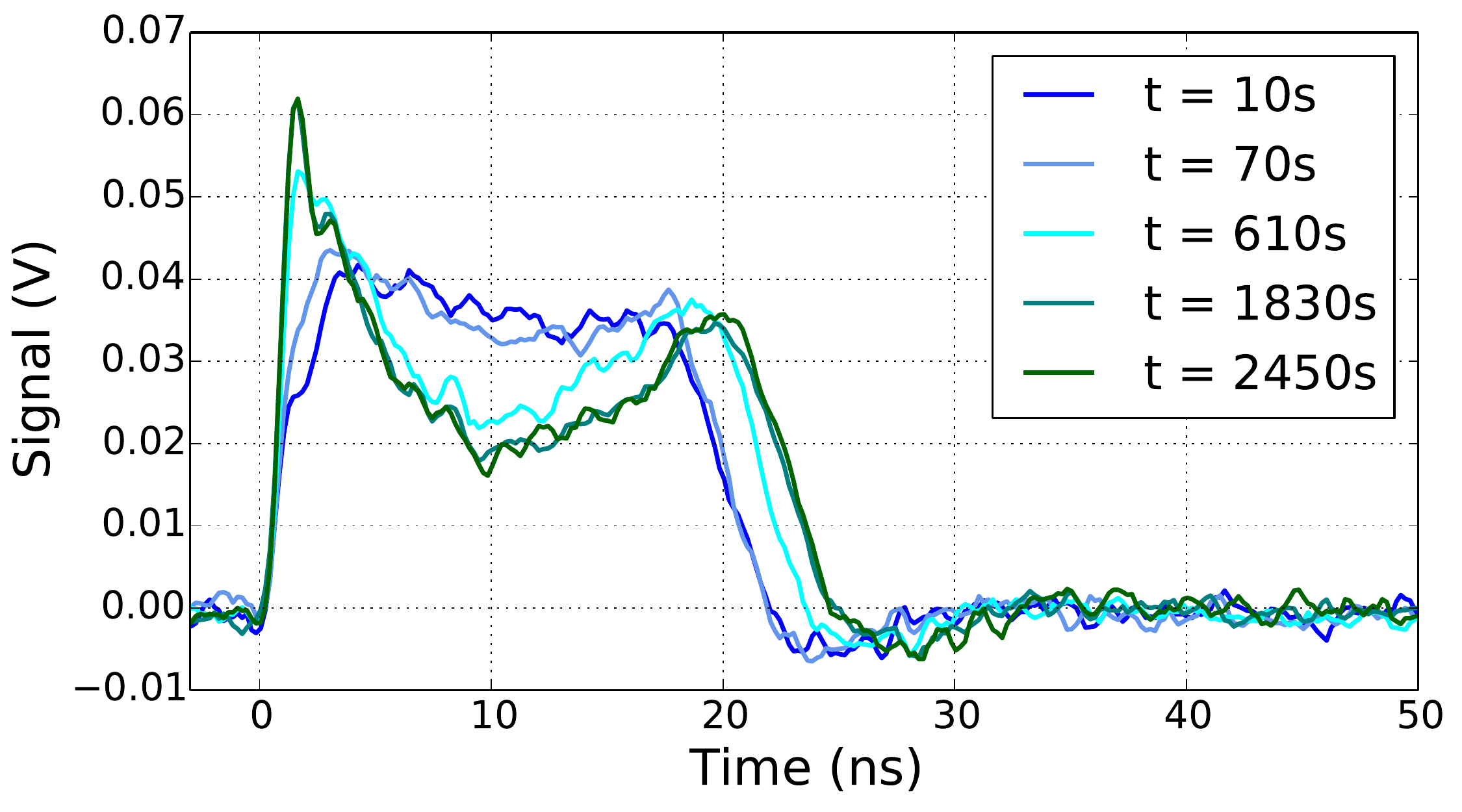}\label{TCT_f0_ele_2}}\hfill
\subfloat[Electron drift: $f_2=3\times 10^{12}\,\rm{cm}^{-2}$]{%
\includegraphics*[width=.3\textwidth]{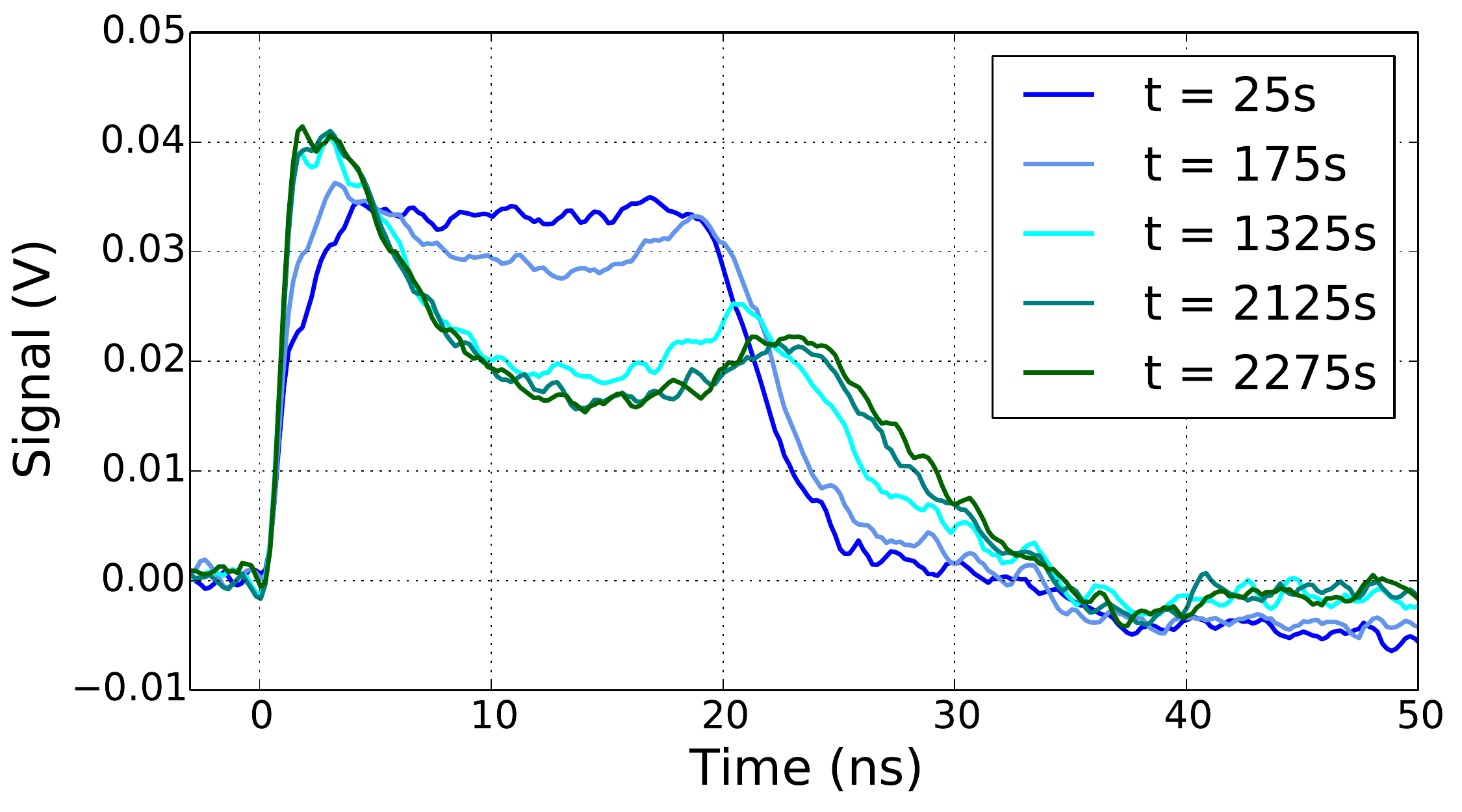}\label{TCT_f0_ele_3}}\hfill
\caption{%
Comparison of the TCT measurements of diamond sample $\#1$ for different radiation damages (from left to right) at a bias voltage of $100\,\rm{V}$ for holes (top row) and electrons (bottom row). The TCT pulses and hence the electrical field is constant over time for the undamaged sensor. Increased radiation damage leads to the diamond polarization and hence to a modified TCT shape corresponding to a modified electrical field configuration. A new stable electrical field configuration is reached for irradiation step $f_1$ after being exposed for $t \geq 610\,\rm{s}$ to the $\beta$ particle source and for irradiation step $f_2$ after $t \geq 2125\,\rm{s}$. With increased radiation damage the effect of the diamond polarization is more significant resulting in a stronger modification of the final TCT pulse shape. The shape of the hole and electron drift is comparable and indicating a symmetrical charge up of the diamond bulk.}
\label{TCT_Polarization}
\end{figure*}

\begin{figure*}%
\subfloat[TCT: hole drift at $100\,\rm{V}$]{%
\includegraphics*[width=.3\textwidth]{pictures/TCT_Hole_Damage2.pdf}\label{TCT_f2_hole_1}}\hfill
\subfloat[TCT: hole drift at $200\,\rm{V}$]{%
\includegraphics*[width=.3\textwidth]{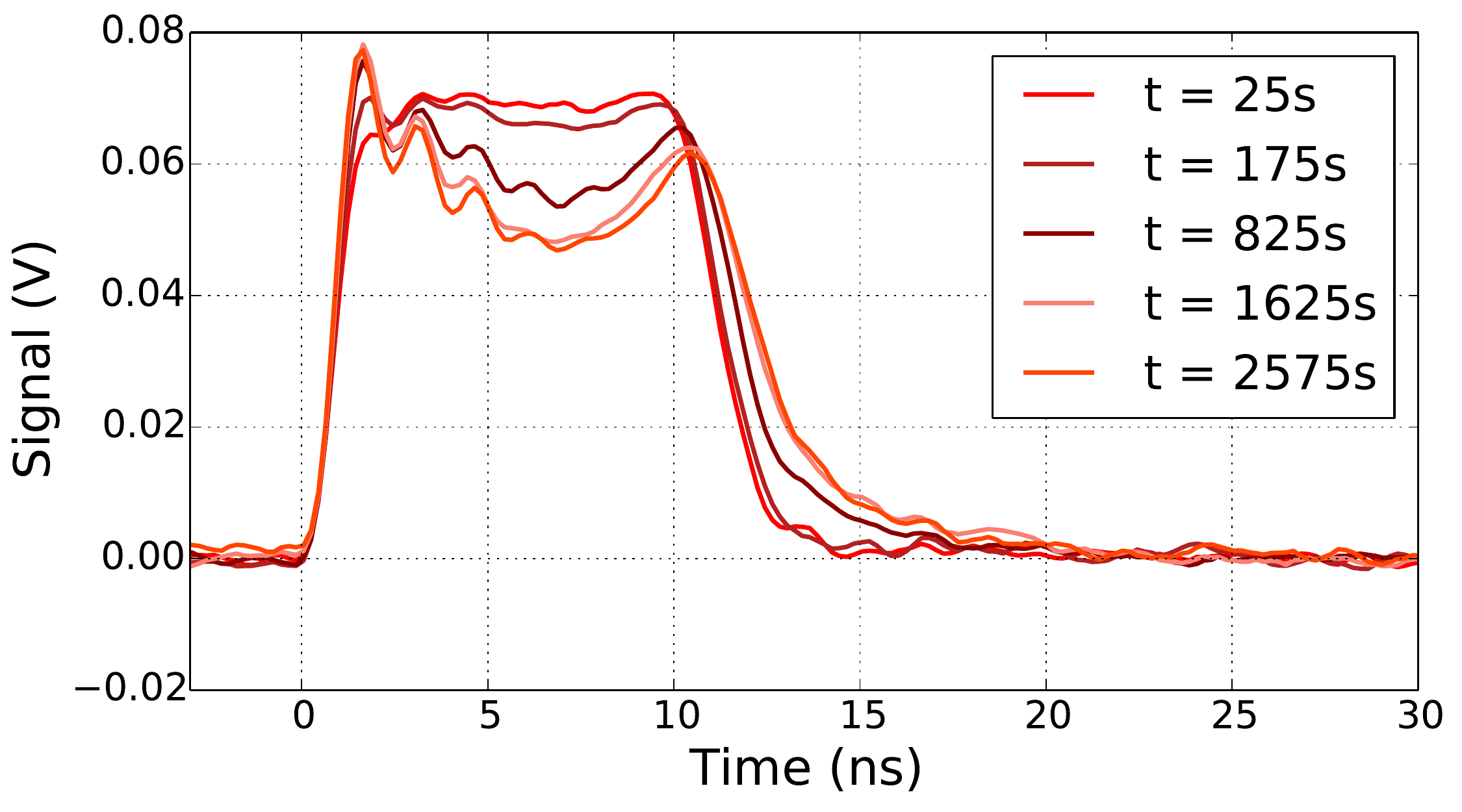}\label{TCT_f2_hole_2}}\hfill
\subfloat[CCE over time at $100\,\rm{V}$ (blue) and $200\,\rm{V}$ (green).]{%
\includegraphics*[width=.3\textwidth]{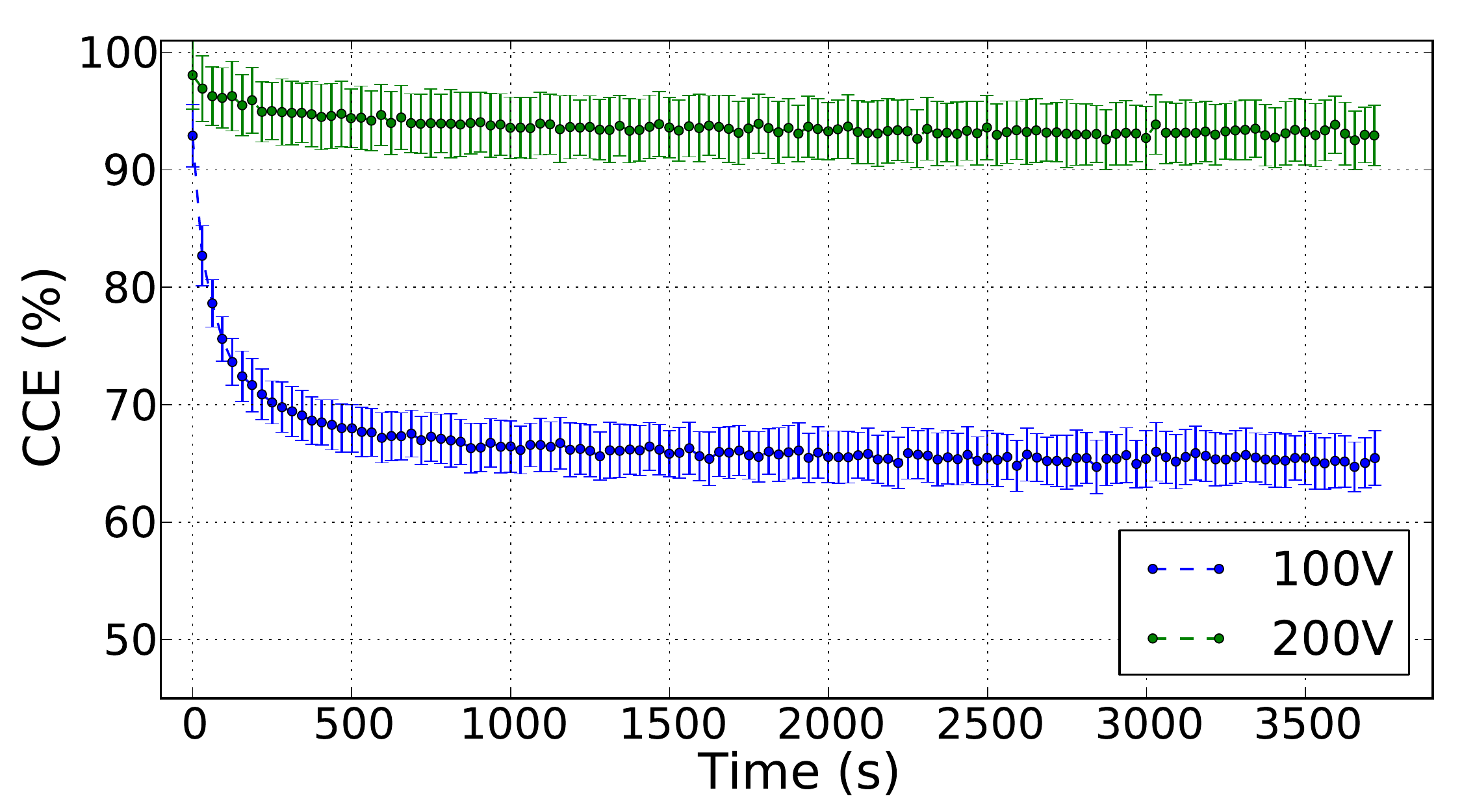}\label{TCT_f2_CCE}}\hfill
\caption{%
Correlation between TCT and CCE measurements at different bias voltages for diamond sample $\#1$ irradiated with $23\,\rm{MeV}$ protons to a total fluence of $f_2=3\times 10^{12}\,\rm{cm}^{-2}$. Increasing the bias voltage leads to a reduction of the diamond polarization and to an increase in the polarized CCE value from $65\,\%$ to $94\,\%$. The TCT and CCE measurements at a bias voltage of $100\,\rm{V}$ show strong changes as function of ionization time, but for an increased voltage of $200\,\rm{V}$ stay almost constant.}
\label{CCE_Polarization}
\end{figure*}

\begin{figure}%
\includegraphics*[width=\linewidth]{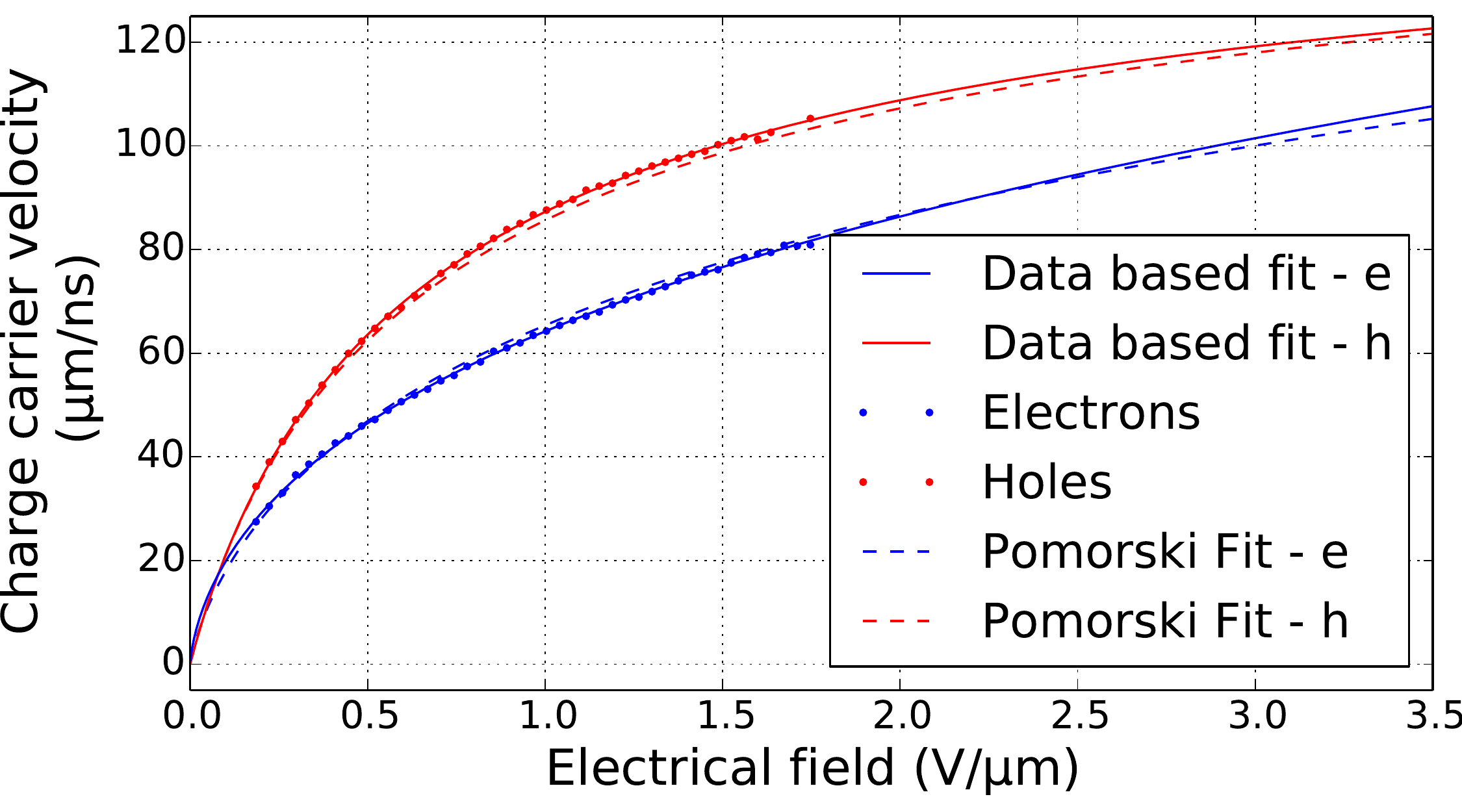}
\caption{%
  Comparison between the experimental measurements of the charge carriers drift velocities (blue and red dots) and literature values (dashed lines) \cite{Pomorski} shows agreement. The best fit to the data is shown as solid lines.}
\label{DriftVelocity}
\end{figure}

\begin{table}
\centering
  \caption{Experimental measured mobility parameters compared to literature values \cite{Pomorski}. The saturation velocity can be calculated using: $v_{\rm{sat}} = E_c/\mu_0$.}
  \begin{tabular}[htbp]{@{}ccccc@{}}
    \hline
    Parameters & $E_c$ & $\mu_0$ & $\beta$ \\
    & ($\rm{kV/cm}$) & ($\rm{cm}^2/\rm{Vs}$)  & \\
    \hline
    $e_{exp}$ & $4.595\pm 0.868$ & $9793\pm 1851$ & $0.30\pm 0.03$ \\
    $h_{exp}$ & $5.787\pm 0.113$ & $2661\pm 44$ & $0.86\pm 0.02$ \\
    $e_{lit}$ & $5.779\pm 0.772$ & $4551\pm 500$ & $0.42\pm 0.01$\\
    $h_{lit}$ & $5.697\pm 0.529$ & $2750\pm 70$ & $0.81\pm 0.01$\\
    \hline
  \end{tabular}
  \label{Drift_parameters_table}
\end{table}

\subsection{Experimental results}\label{Experimental_results_chapter}
The analysis presented focuses on one diamond sample ($\#1$) out of the four irradiated diamond samples. This diamond sample was irradiated in two steps with $23\,\rm{MeV}$ protons. The first irradiation step was $f_1=1\times 10^{12}\,\rm{cm}^{-2}$ and the second step  $f=2\times 10^{12}\,\rm{cm}^{-2}$ to a total fluency of $f_2=3\times 10^{12}\,\rm{cm}^{-2}$. 
For the unirradiated diamond sample the TCT pulses show the expected rectangular shape corresponding to a constant electrical field, see Figs.~\ref{TCT_f0_hole_1} and \ref{TCT_f0_ele_1}. The constant $\beta$ particle rate causes no effect and the TCT pulses remain constant over the entire measurement time. The CCE was measured as well and remained stable at $95\,\%$.

After the first irradiation step $f_1$ the TCT pulses are no longer stable over time. A rectangular TCT pulse shape can however still be obtained immediately after the start of the measurement ($t=15\,\rm{s}$ in Fig.~\ref{TCT_f0_hole_2} and $t=10\,\rm{s}$ in \ref{TCT_f0_ele_2}). A new stable configuration is reached after $t=615\,\rm{s}$ in Fig.~\ref{TCT_f0_hole_2} and $t=610\,\rm{s}$ in Fig.~\ref{TCT_f0_ele_2}, when the TCT pulse shapes are not changing anymore. The TCT pulses of both charge carriers show a symmetrical two-peak structure, indicating a local minimum in the electrical field central in the diamond bulk. 

After the second irradiation step $f_2$ the TCT pulse modification is more pronounced, as shown in Figs.~\ref{TCT_f0_hole_3} and \ref{TCT_f0_ele_3}. The time to reach a new stable configuration is now increased to $t=2125\,\rm{s}$. The TCT pulse shapes of both charge carriers are still similar, indicating a symmetrical charge up of the diamond bulk with a local electrical field minimum in the center.

The correlation between the TCT pulse modification and the CCE for the diamond sample ($\#1$) at the irradiation step $f_2$ for different bias voltages is shown in Fig.~\ref{CCE_Polarization}. At the bias voltage of $100\,\rm{V}$ the strong modification of the TCT pulse (electrical field), see Fig.~\ref{TCT_f2_hole_1}, is also visible in the drop in CCE from $93\,\%$ to $65\,\%$, see Fig.~\ref{TCT_f2_CCE}. A stable CCE value is reached after approximately $t=2000\,\rm{s}$ comparable to the TCT measurement reaching a stable TCT configuration after $t=2125\,\rm{s}$.

An increase in bias voltage to $200\,\rm{V}$ results in a less modified TCT shape and in a higher CCE as shown in Figs.~\ref{TCT_f2_hole_2} and \ref{TCT_f2_CCE}. This is expected, since the electric field is not so strongly reduced by the internal field from the polarization.

\paragraph{Mobility of charge carriers}
The mobility and drift parameters of the charge carriers inside the diamond are crucial for a correct TCT pulse simulation. The drift velocity of the charge carriers can be calculated for different bias voltages, by measuring the FWHM of the TCT pulse and using Eq.~\ref{DriftVelocity_eq}:
\begin{equation}
\label{DriftVelocity_eq}
v_{\rm{drift}} = \frac{d}{\tau_{\rm{FWHM}}},
\end{equation}
where $d$ is the thickness of the diamond. The measured drift velocities for bias voltages between $70\,\rm{V}$ and $1000\,\rm{V}$ are in agreement with the literature values \cite{Pomorski} shown in Fig.~\ref{DriftVelocity}.

Caughey and Thomas \cite{Caughey67} found that the drift velocities can be well described by the empirical formula:
\begin{equation}
\label{DriftVelocity_eq_2}
v_{\rm{drift}} = v_{\rm{sat}}\frac{E/E_c}{(1+(E/E_c)^{\beta})^{1/\beta}},
\end{equation}
with
\begin{equation}
\label{DriftVelocity_eq_3}
E = v_{\rm{drift}}/\mu_{\rm{0}}.
\end{equation}
The fitted mobility parameters $\mu_0, \beta $ and $E_c$ are compared with the literature parameters in Table~\ref{Drift_parameters_table}.

\section{Quasi 3D-simulation of diamond polarization}
In order to gain a quantitative understanding of the diamond polarization the diamond sensor was modeled with the software SILVACO TCAD \cite{Silvaco}. Besides the electrical properties, like e.g. band gap or mobility parameters, radiation induced lattice defects can be taken into account by introducing effective deep traps acting as recombination centers. These effective recombination centers were found by optimizing the simulation of TCT pulses to match the experimental data. Beside TCT pulses, the CCE of the detector can be obtained by simulating a MIP particle hit.

\begin{figure}%
\includegraphics*[width=\linewidth]{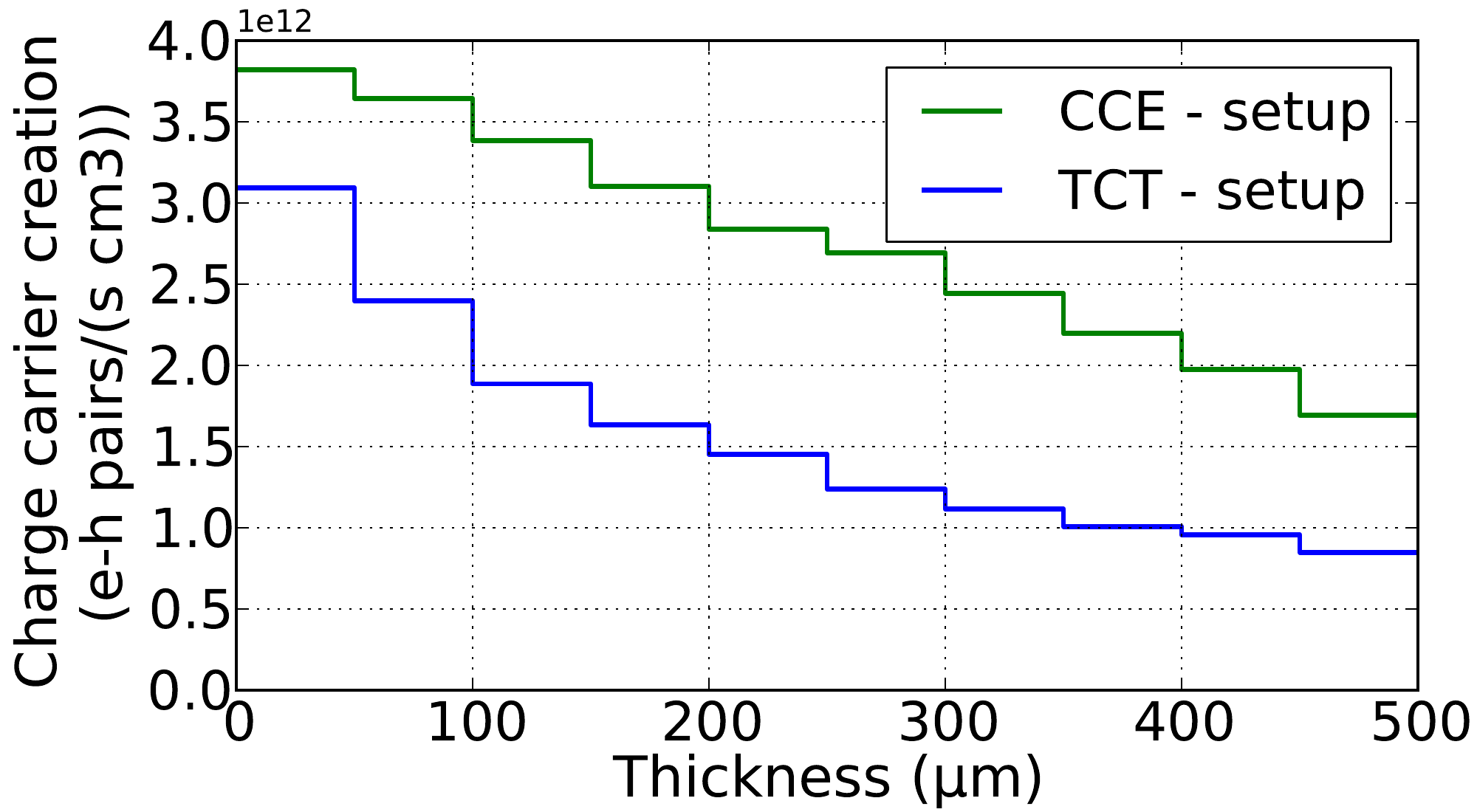}
\caption{%
  The MIP particle rate created by the $^{90}Sr$ sources is simulated in FLUKA for both measurement setups. The activity of both $\beta$ particle sources is comparable (TCT: $32.2\,\rm{MBq}$ and CCE: $33.4\,\rm{MBq}$). However, the different geometries in the set\-ups lead to an increased charge carrier creation inside the diamond bulk for the CCE setup.}
\label{ChargeCarriers}
\end{figure}

\begin{figure}%
\includegraphics*[width=\linewidth]{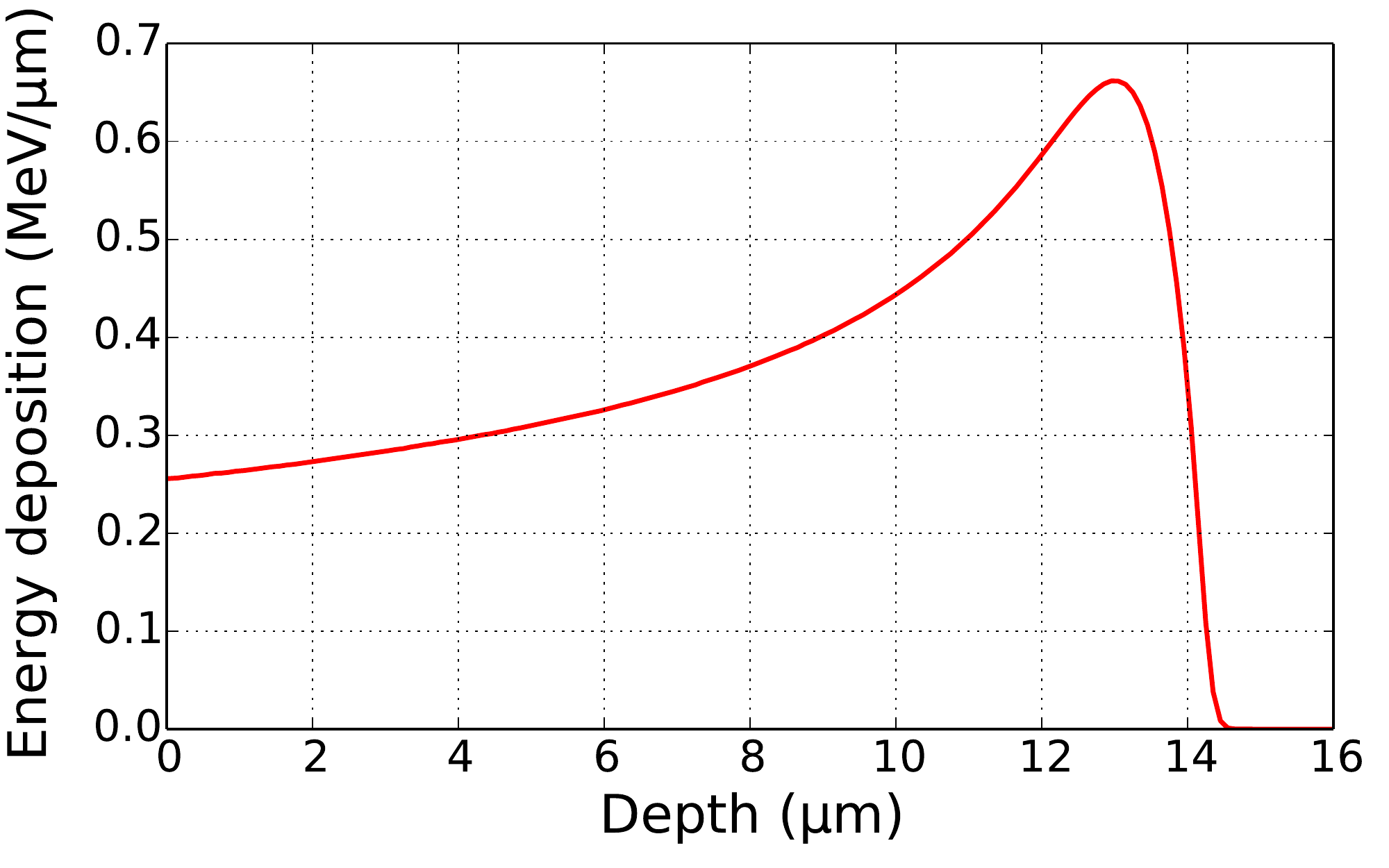}
\caption{%
  FLUKA simulation of the energy deposition of an $\alpha$ particle ($5.5\,\rm{MeV}$) hit in diamond.}
\label{AlphaInDiamond}
\end{figure}

\begin{figure}%
\includegraphics*[width=\linewidth]{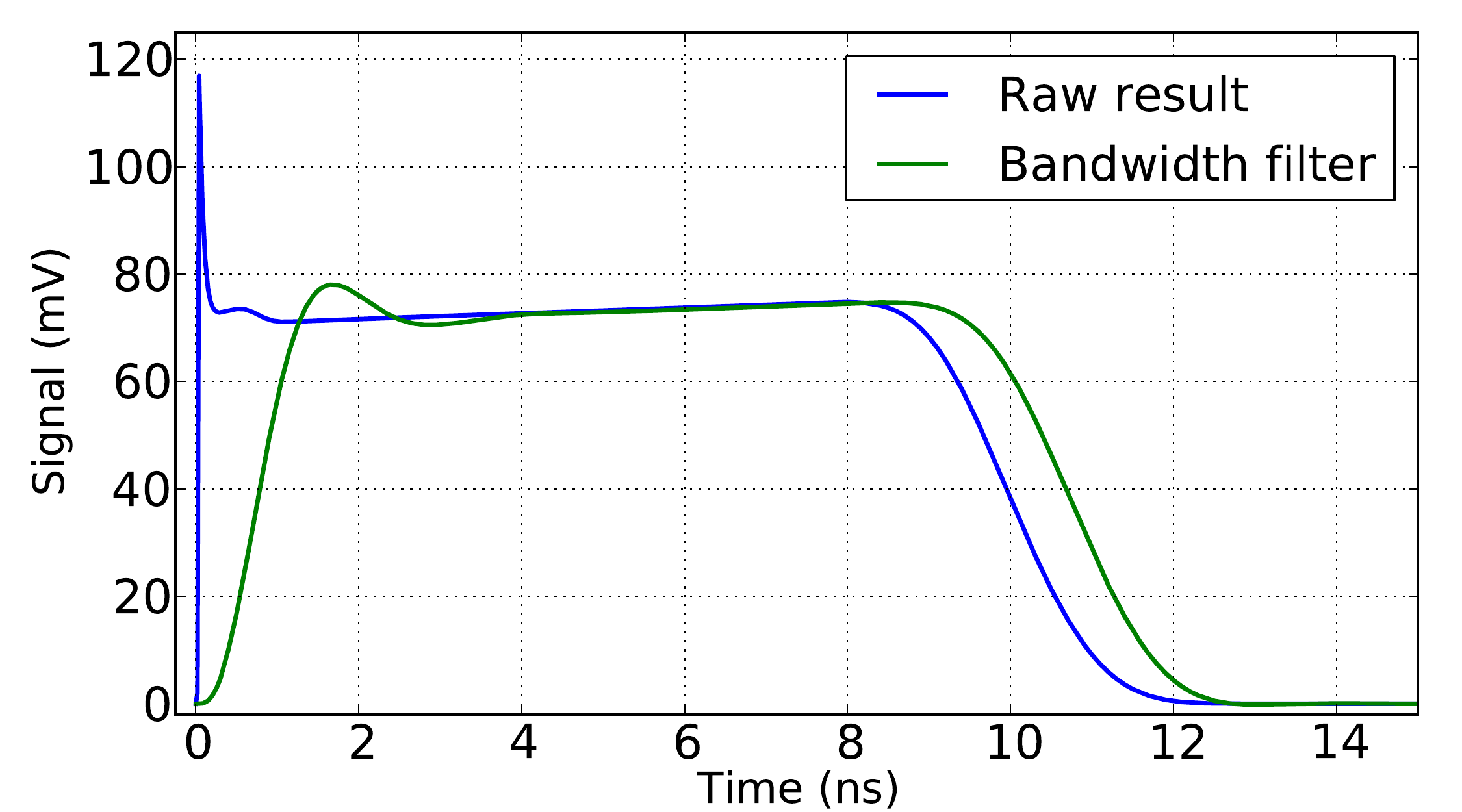}
\caption{%
  The raw TCT signal calculated by the simulation (blue) is manipulated with a bandwidth filter to simulate the TCT hardware having a limited bandwidth of $1\,\rm{GHz}$ (see section: \ref{TCT_setup_chapter}). The unfiltered simulation result shows a narrow peak at the beginning of the TCT pulse, which is caused by the drift of both charge carriers immediately after the alpha particle hit.}
\label{Limited_Bandwidth}
\end{figure}

\begin{figure}%
\includegraphics*[width=\linewidth]{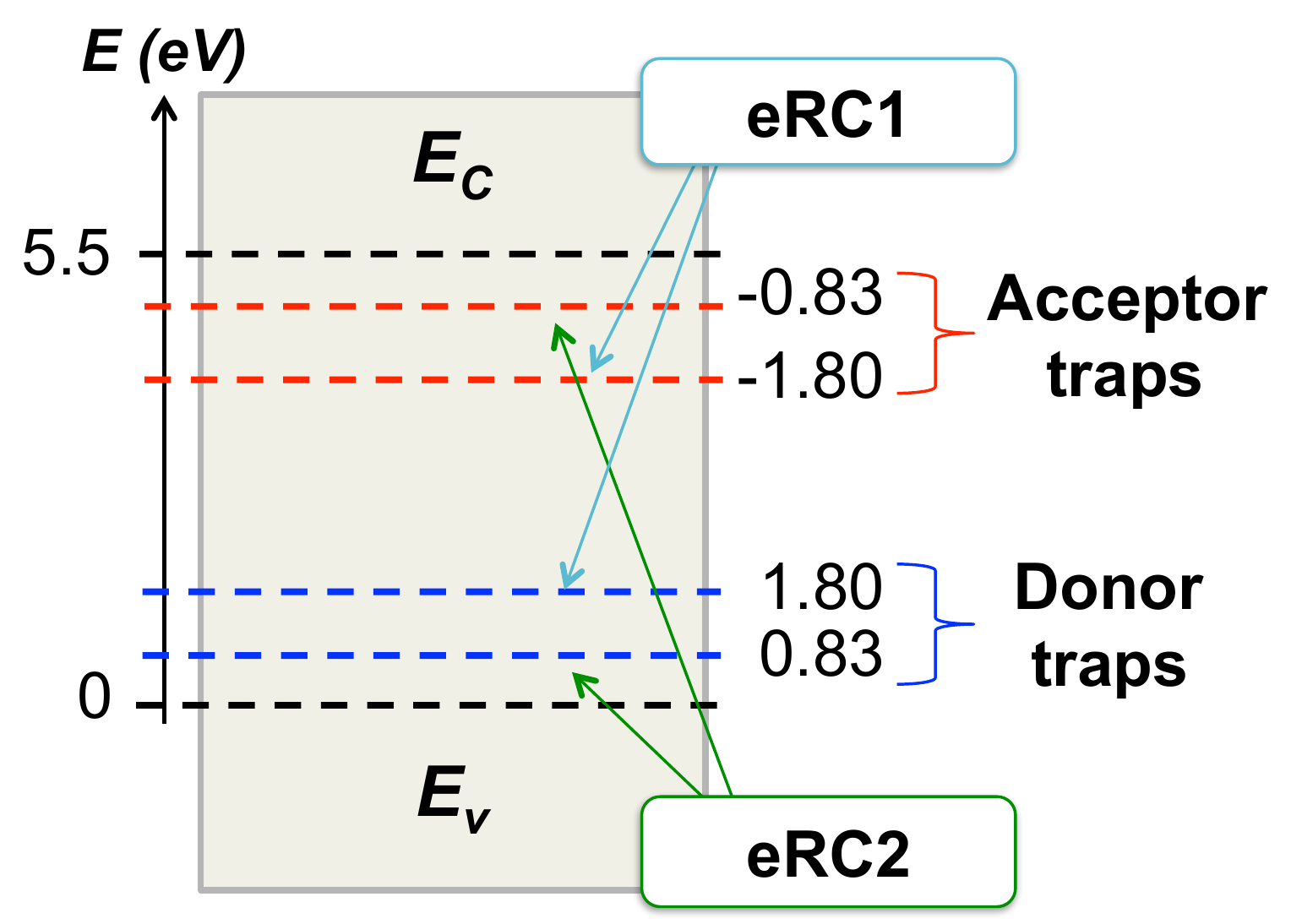}
\caption{%
  Location of the effective recombination centers $eRC1/2$ in the band structure.}
\label{Traps}
\end{figure}

\subsection{Description of simulation}\label{simulation_desc_chapter}
The electrical diamond properties are simulated numerically in a 2 dimensional rotational symmetrical system (quasi 3D), allowing extrapolating quantities like the charge density into a three dimensional sphere. This approach combines an efficient use of computing resources with a correct description of all physical quantities.

The experimental determined mobility pa\-ra\-meters (section \ref{Experimental_results_chapter}) for the hole drift and the literature parameters for the electron drift are used in the simulation. The diamond polarization can lead to extreme electrical fields ($\overrightarrow{E}<0.1\,\rm{V/cm}$ or $\overrightarrow{E}>10\,\rm{V/cm}$), where the experimental determined electron parameterization could fail.

The energy deposition of the $\alpha$ and $\beta$ patricles in the diamond bulk during the TCT and CCE measurements were simulated with the publicly available software FLUKA \cite{FLUKA1,FLUKA2}. The simulation includes the geometry of the setups and the different collimation of the sources (see Figs.~\ref{ChargeCarriers} and \ref{AlphaInDiamond}).

Besides the geometrical properties of the measurement setup, the hardware limitation of the scope and the amplifier are taken into account. Therefore the raw TCT signal calculated by the simulation is subjected to a bandwidth filter (see Fig.~\ref{Limited_Bandwidth}) to simulate the TCT hardware having a limited bandwidth of $1\,\rm{GHz}$ (see Sect: \ref{TCT_setup_chapter}). The unfiltered simulation result shows a narrow peak at the beginning of the TCT pulse, which is caused by the drift of both charge carriers in opposite directions immediately after the alpha particle hit. The small penetration depth $d_{\rm{alpha}}\leq15\,\rm{\mu m}$ (see Fig.~\ref{AlphaInDiamond}) explains the short duration of the peak ($t<0.2\,\rm{ns}$).

\begin{table}
\centering
  \caption{Physical parameters of the effective recombination centers. Both effective recombination centers $eRC1/2$ are present as acceptor- and donor-like traps.}
  \begin{tabular}[htbp]{@{}lcc@{}}
    \hline
    Trap name & $eRC1$ & $eRC2$ \\
    \hline
    Energy level $E_t$ (eV)  &  1.8 & 0.83 \\
    Density $\rho$ ($\rm{cm}^{-3}$)  & 9.98e11 & 1.44e12 \\
    Cross section $\sigma_e$ ($\rm{cm}^2$) & 7e-15	& 2e-14 \\
    Cross section $\sigma_h$ ($\rm{cm}^2$) & 7e-15	& 1e-14 \\
    \hline
  \end{tabular}
  \label{Traps_table}
\end{table}

\begin{figure}%
\includegraphics*[width=\linewidth]{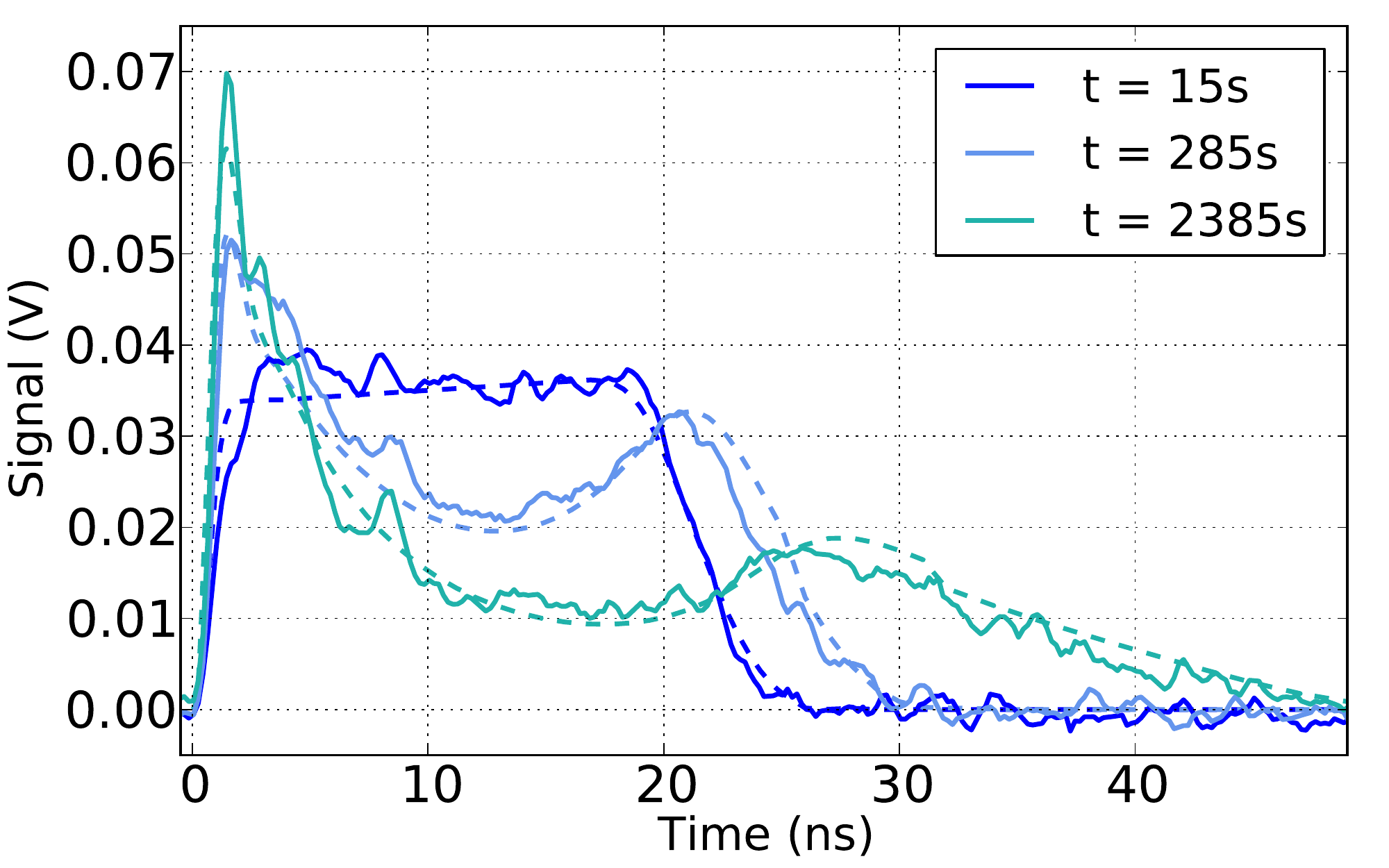}
\caption{%
  Comparison of the TCT pulse deformation for an electron drift over time between simulation (dashed) and measurement (solid). The evolution in time of the TCT pulse is in agreement with the experimental data.}
\label{TCT_100V_ele_full}
\end{figure}

\begin{figure*}[bth]%
\subfloat[TCT: hole drift at $100\,\rm{V}$]{%
\includegraphics*[width=.3\textwidth]{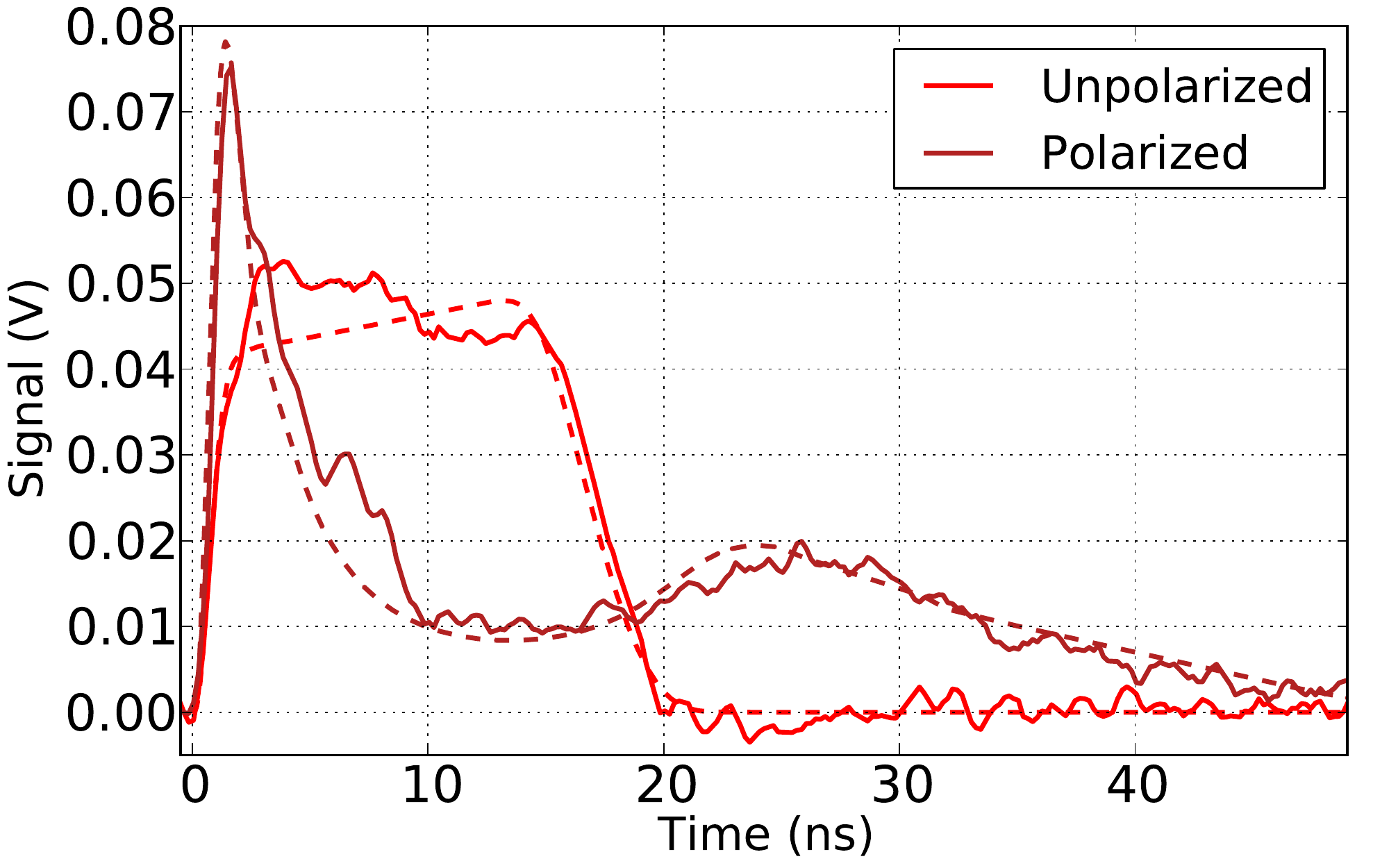}}\hfill
\subfloat[TCT: electron drift at $100\,\rm{V}$]{%
\includegraphics*[width=.3\textwidth]{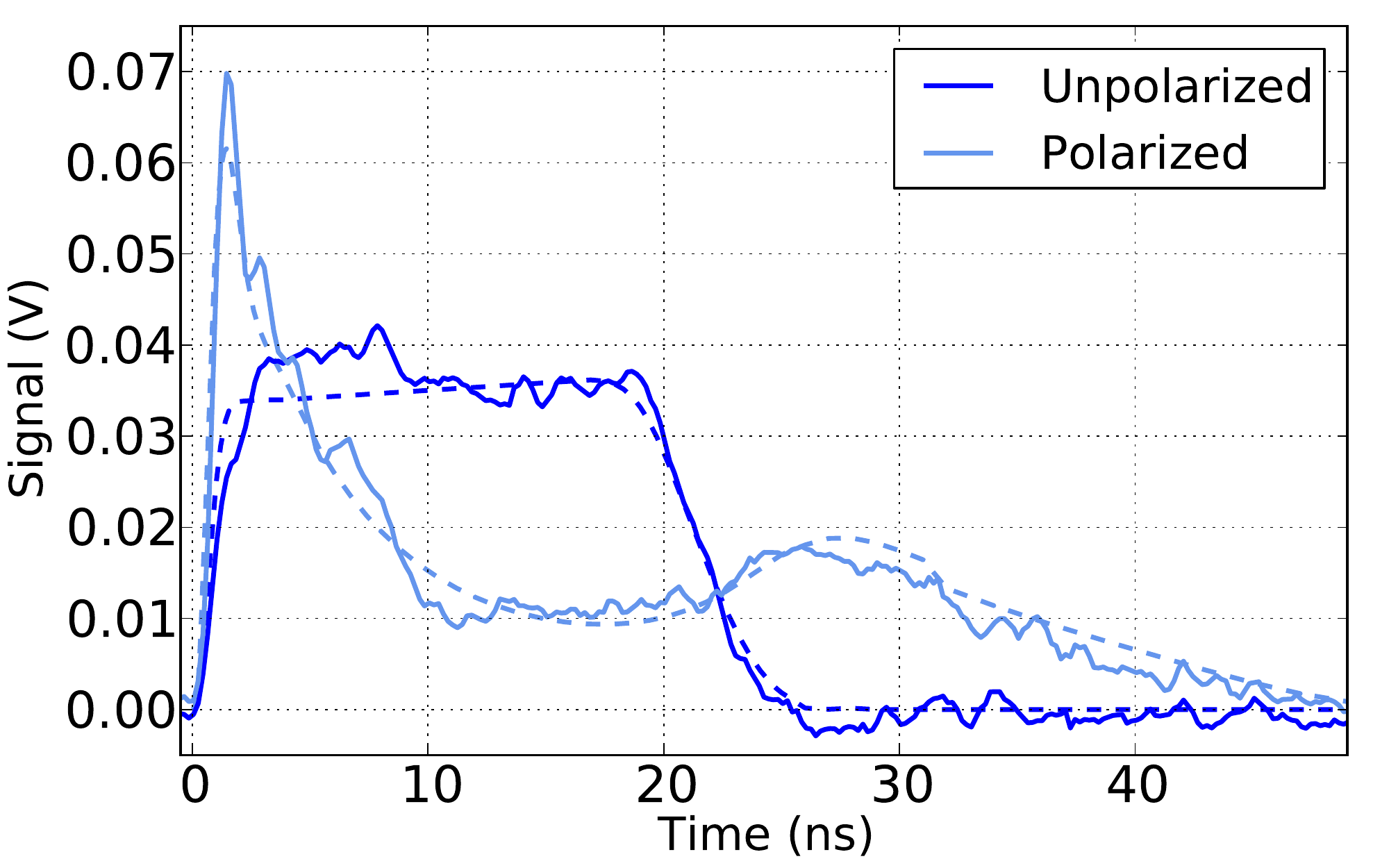}}\hfill
\subfloat[CCE over time: $100\,\rm{V}$]{%
\includegraphics*[width=.3\textwidth]{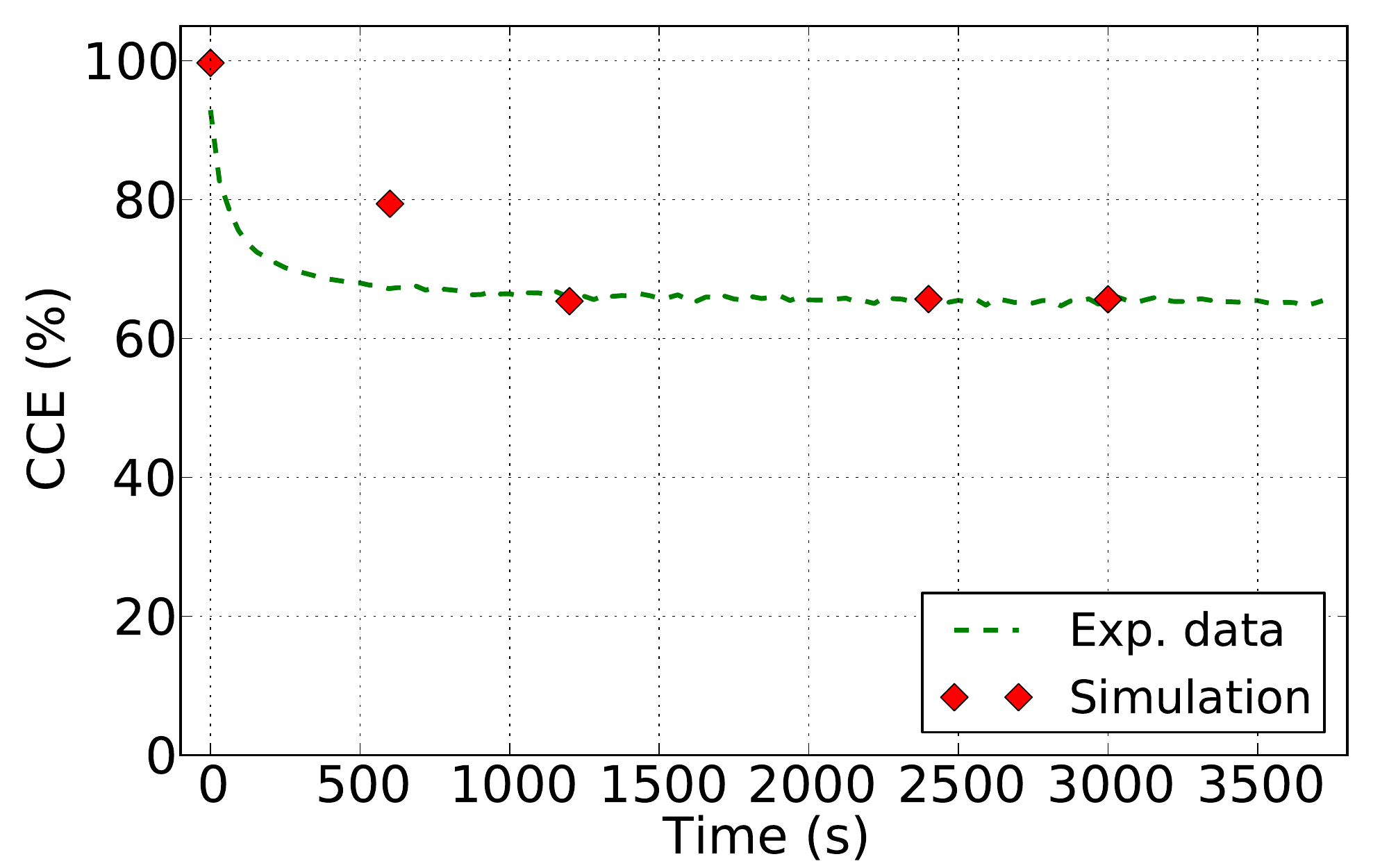}}\hfill
\subfloat[TCT: hole drift at $200\,\rm{V}$]{%
\includegraphics*[width=.3\textwidth]{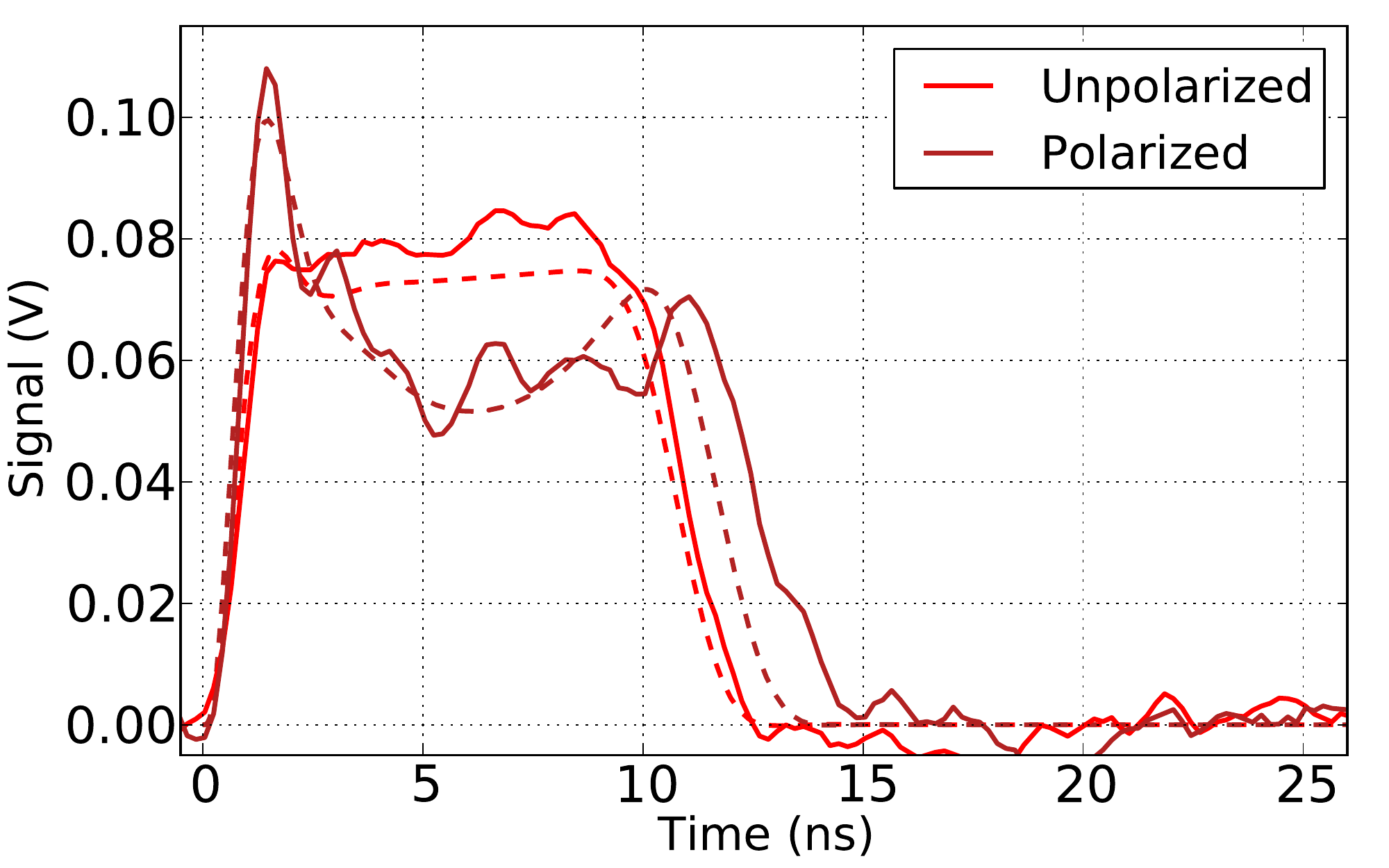}}\hfill
\subfloat[TCT: electron drift at $200\,\rm{V}$]{%
\includegraphics*[width=.3\textwidth]{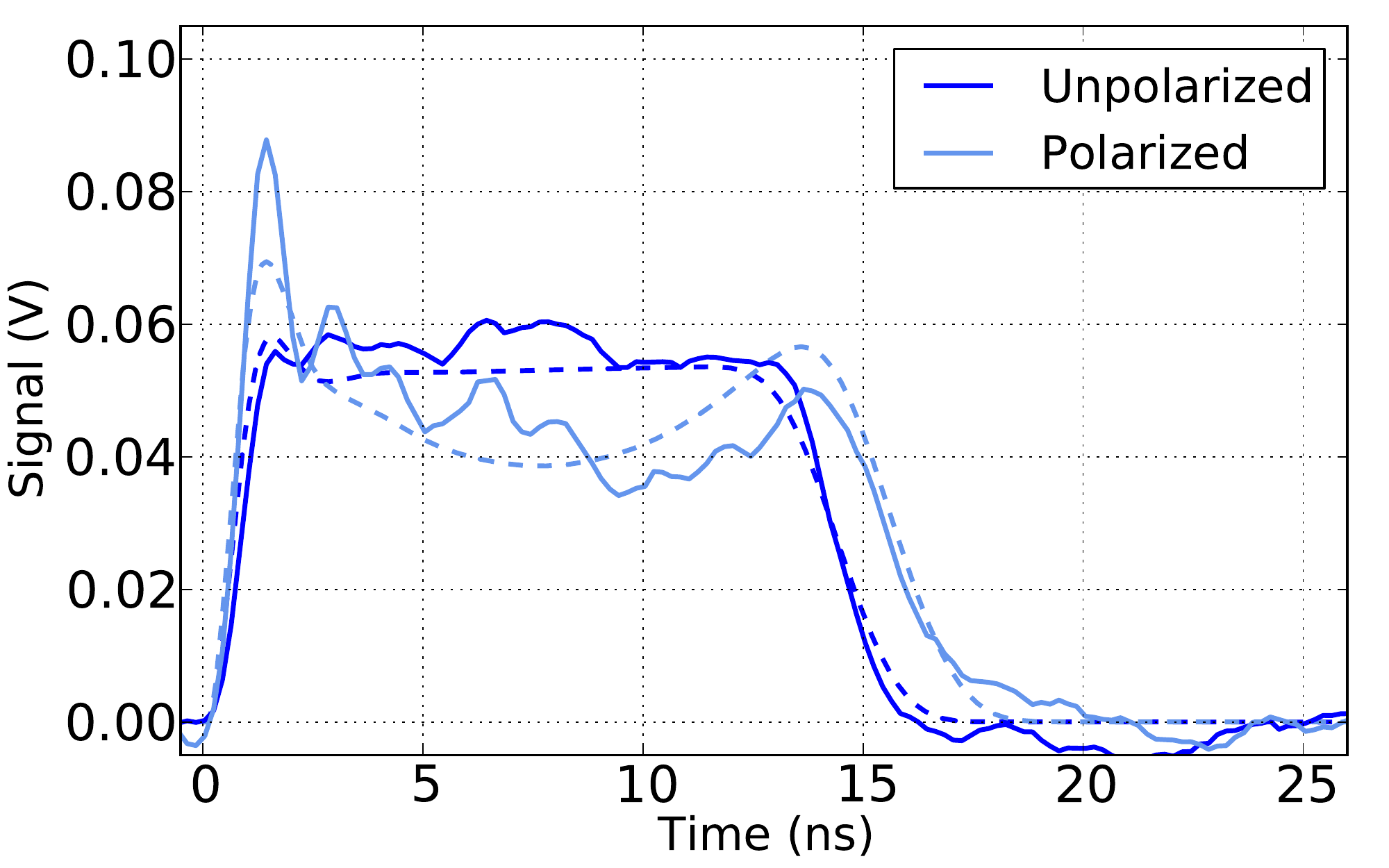}}\hfill
\subfloat[CCE over time: $200\,\rm{V}$]{%
\includegraphics*[width=.3\textwidth]{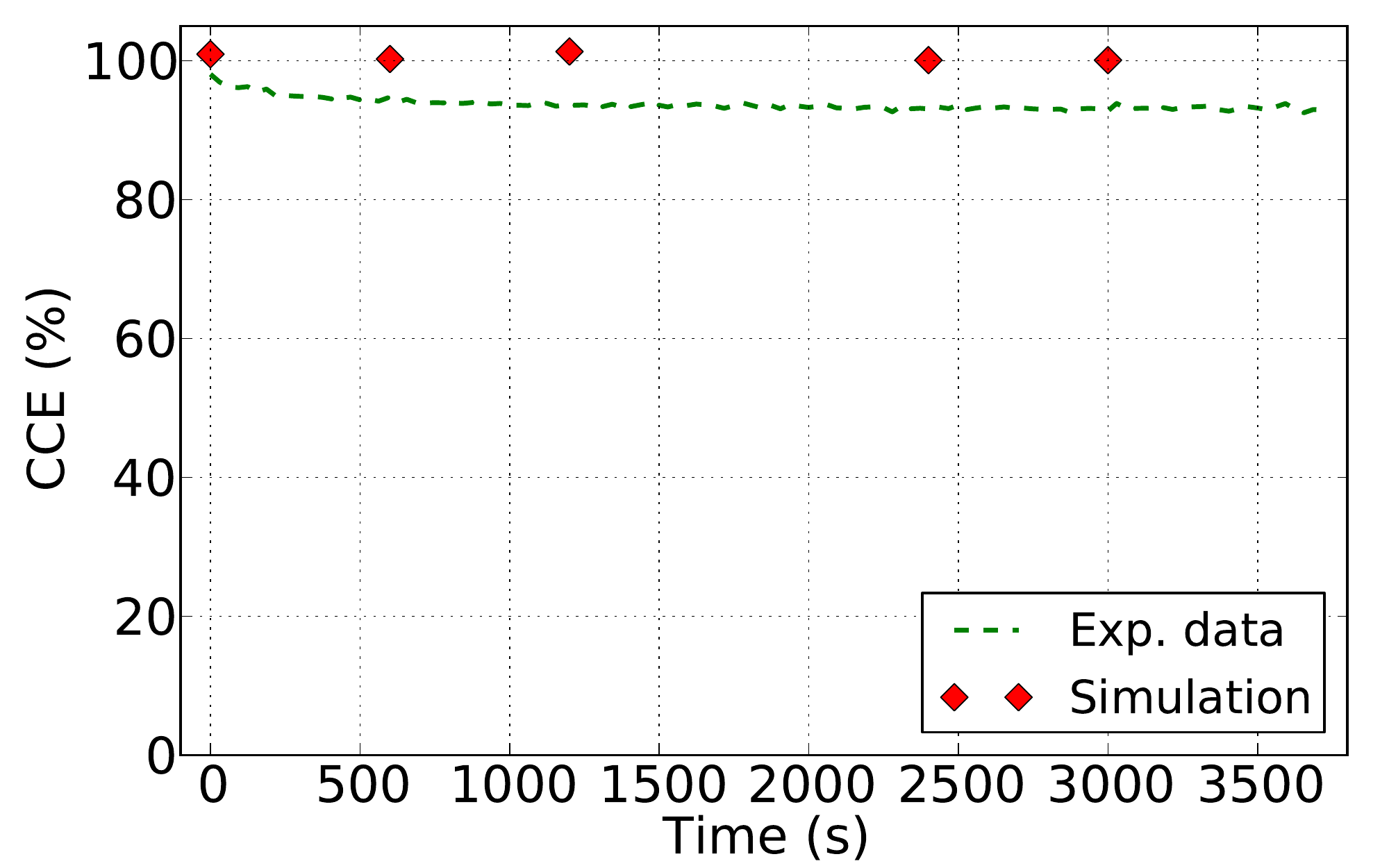}}\hfill
\subfloat[TCT: hole drift at $400\,\rm{V}$]{%
\includegraphics*[width=.3\textwidth]{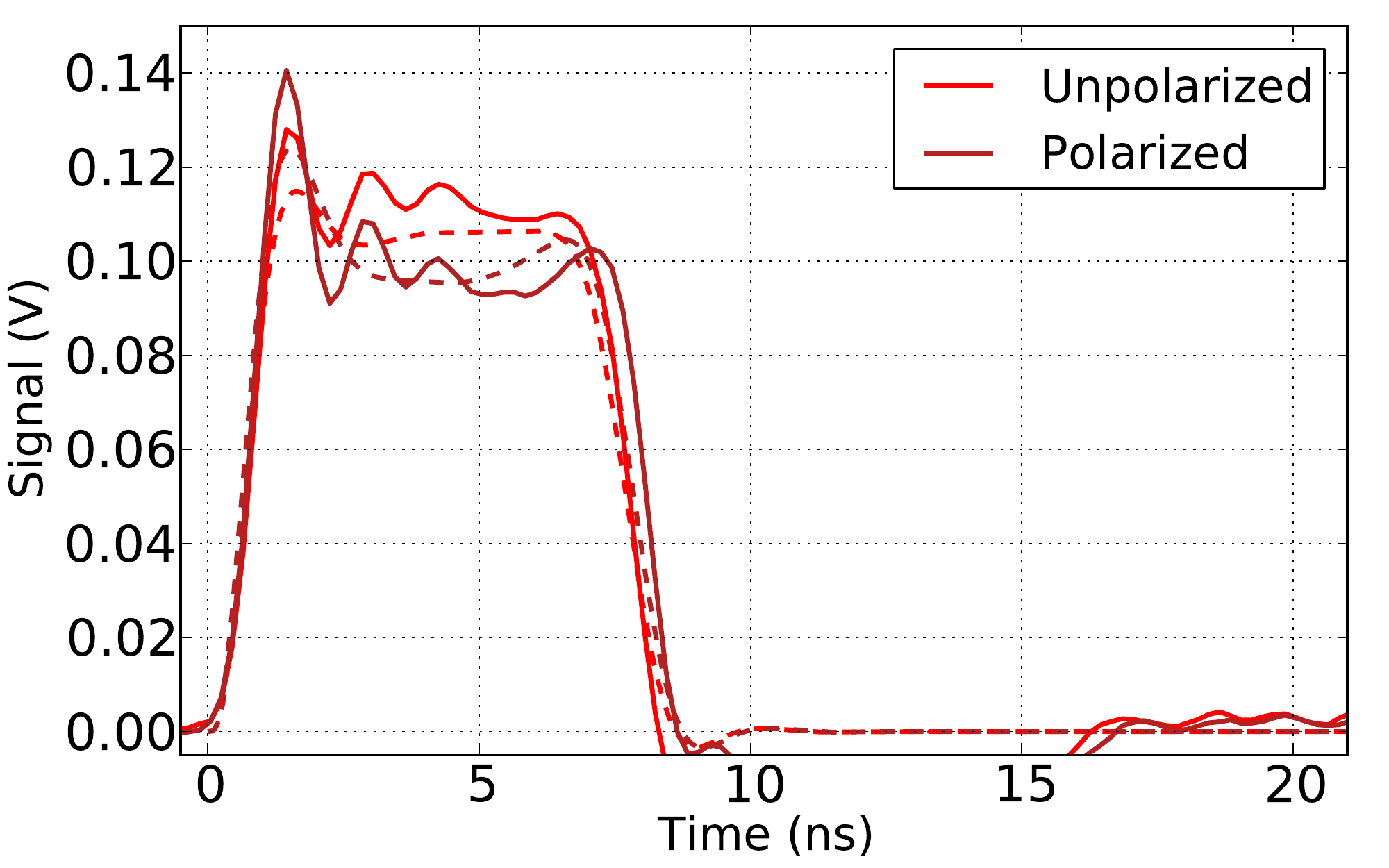}}\hfill
\subfloat[TCT: electron drift at $400\,\rm{V}$]{%
\includegraphics*[width=.3\textwidth]{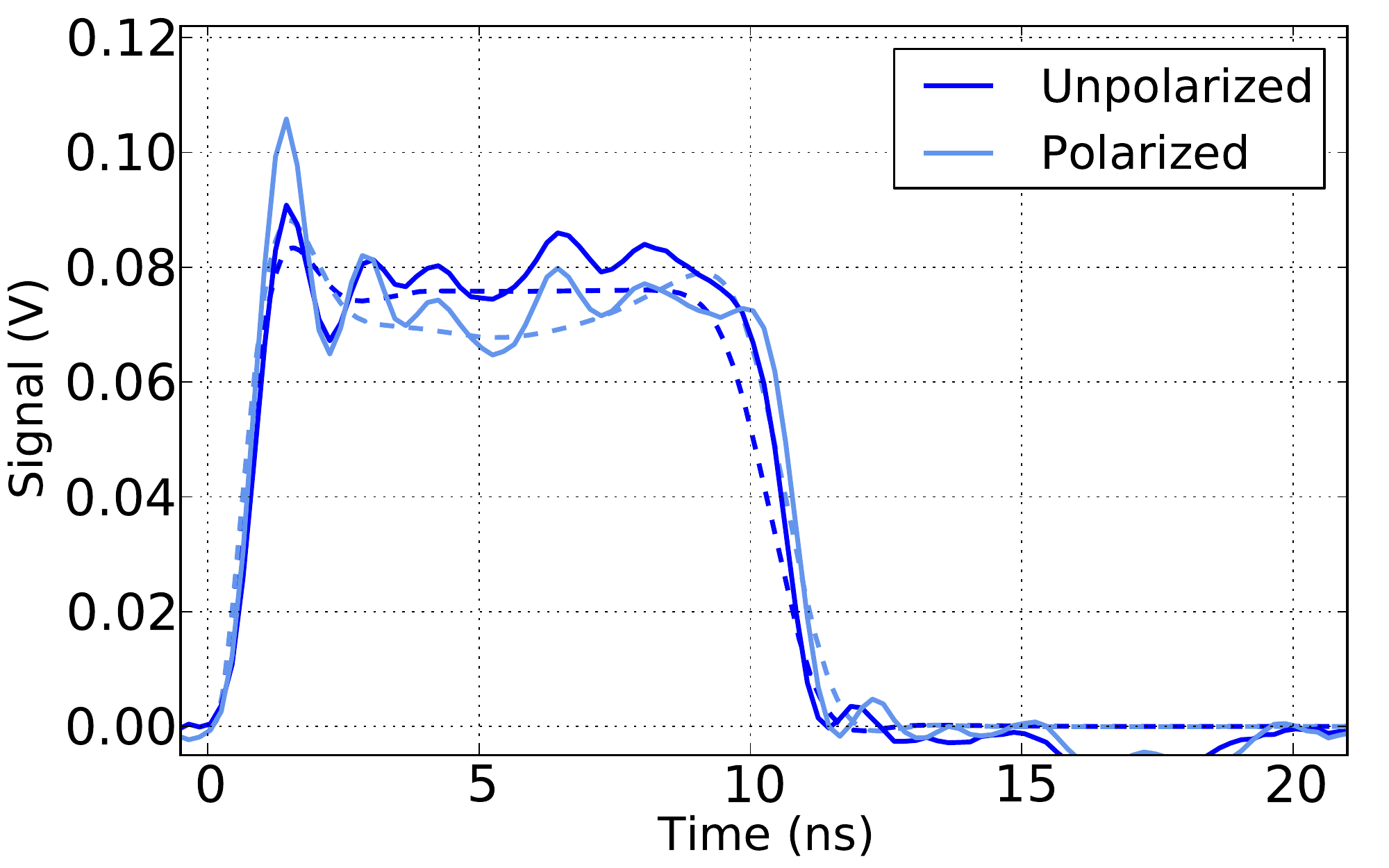}}\hfill
\subfloat[CCE over time: $400\,\rm{V}$]{%
\includegraphics*[width=.3\textwidth]{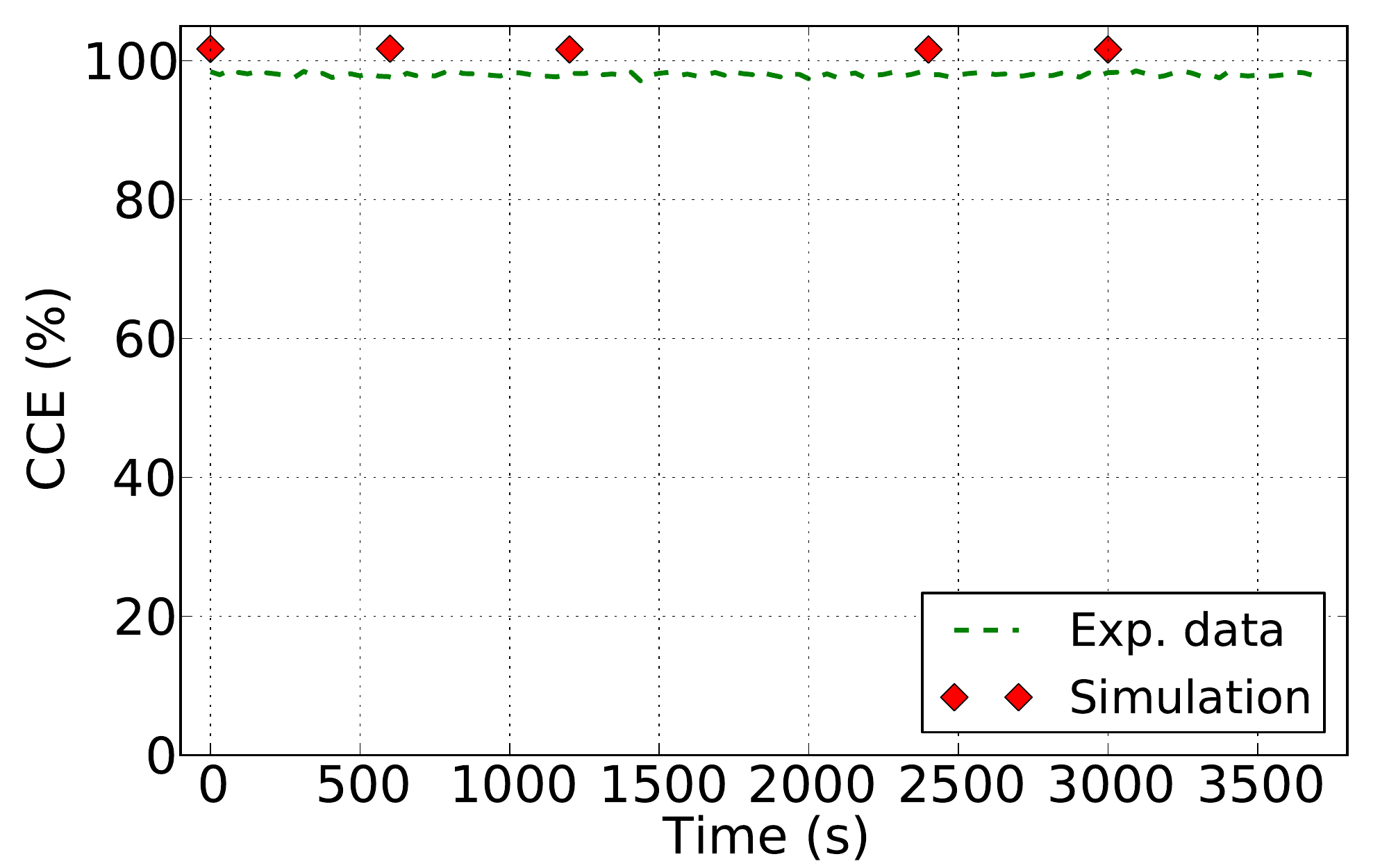}}\hfill
\caption{%
Comparison of the TCT pulse modification between simulation (dashed) and experimental data (solid) for different bias voltages ($100\,\rm{V}$, $200\,\rm{V}$ and $400\,\rm{V}$ from top to bottom) for a diamond sensor irradiated up to a total fluency $f_2=3\times 10^{12}\,\rm{cm}^{-2}$. The CCE simulation results (red stars) are compared to the experimental data (green dashed) on the right side.}
\label{Simulation_100V}
\end{figure*}

\subsection{Effective recombination center model}
The irradiation creates a plutoria of defects, which can trap charges (electrons or holes) or stay in an ionized state. Since the energy levels of the defects are poorly known, we introduced four effective deep traps acting as recombination centers to approximate the many different defects in reality. The energy levels and cross sections of such a simple model are obtained by fitting the evolution of the TCT pulse shapes in time. The positions of the effective recombination centers are illustrated in Fig.~\ref{Traps} and the physical parameters are given in Table~\ref{Traps_table}.

Beside the creation of space charge the physical para\-meters of the defects are influencing recombination processes, especially the Shockley-Read-Hall (SRH) recombination of charge carriers by the release of energy in form of a phonon \cite{SRH1952}. The recombination rate $R_{D}$ is described by the Shockley-Read-Hall law \cite{ATALASmanual}:
\begin{multline}
R_{D}=(eh-n^2_{ie}) \times \biggl[ \tau_{e}\left(h+n_{ie}exp\left(\frac{E_{i}-E_{t}}{kT_L}\right)\right) \\ +\tau_{h}\left(e+n_{ie}exp\left(\frac{E_{t}-E_{i}}{kT_L}\right)\right) \biggr]^{-1},
\label{SRH_model}
\end{multline}
where the electron/hole lifetimes $\tau_{e,h}$ are related to the carrier capture cross sections $\sigma_{e,h}$ through the equation:
\begin{equation}
\tau_{e,h}=\frac{1}{\sigma_{e,h}\cdot v_{e,h}\cdot \rho}.
\label{lifetime_CS}
\end{equation}
$E_i$ is the intrinsic Fermi level position, $E_t$ is the trap energy level and $v_{e,h}$ is the thermal velocity of the charge carriers. Hence the trap parameters determine the electron/hole lifetime and influence directly the amount of charge carrier recombination.

\begin{figure*}[bth]%
\subfloat[Electrical field at $100\,\rm{V}$.]{%
\includegraphics*[width=.3\textwidth]{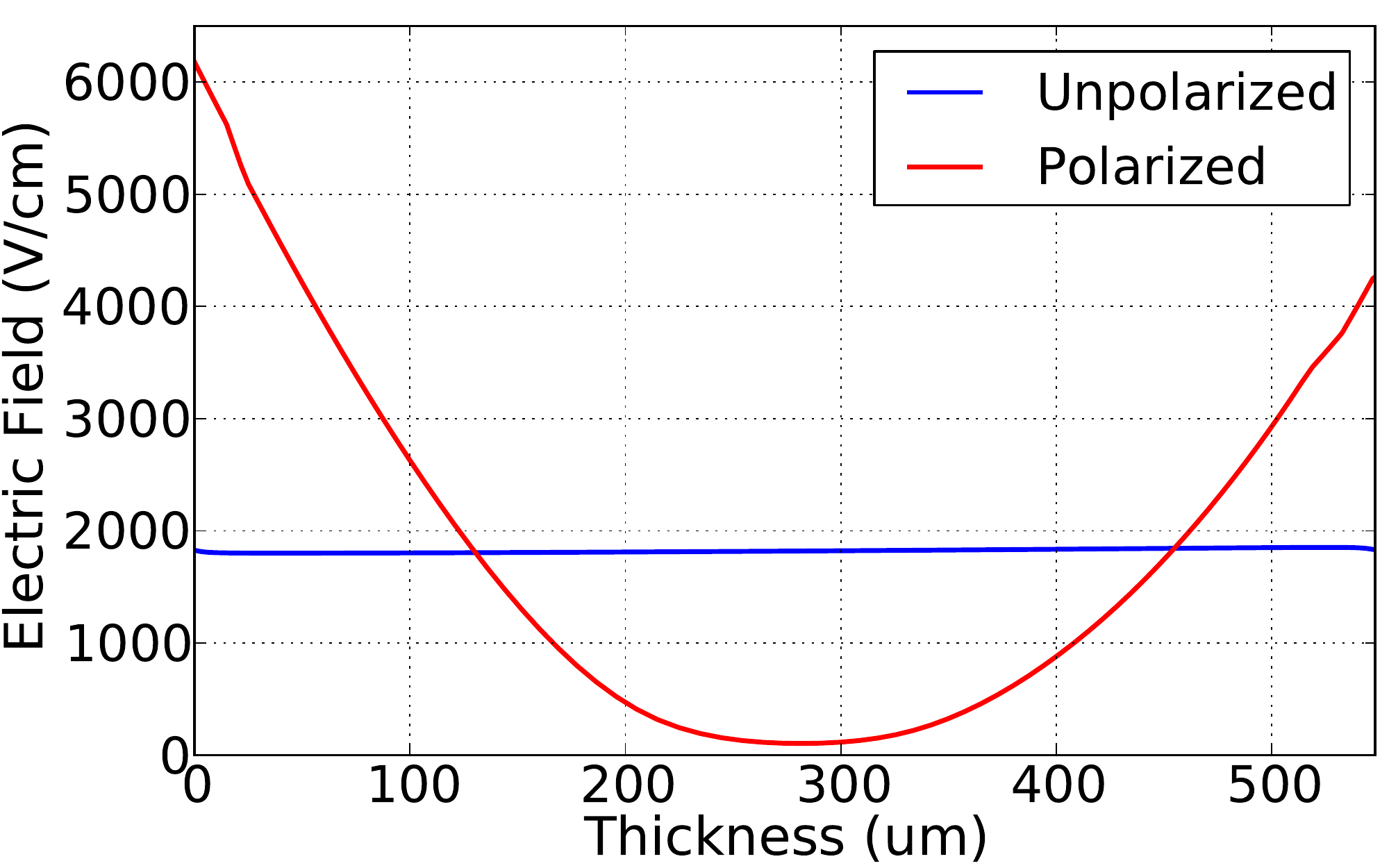}}\hfill
\subfloat[Recombination rate at $100\,\rm{V}$.]{%
\includegraphics*[width=.3\textwidth]{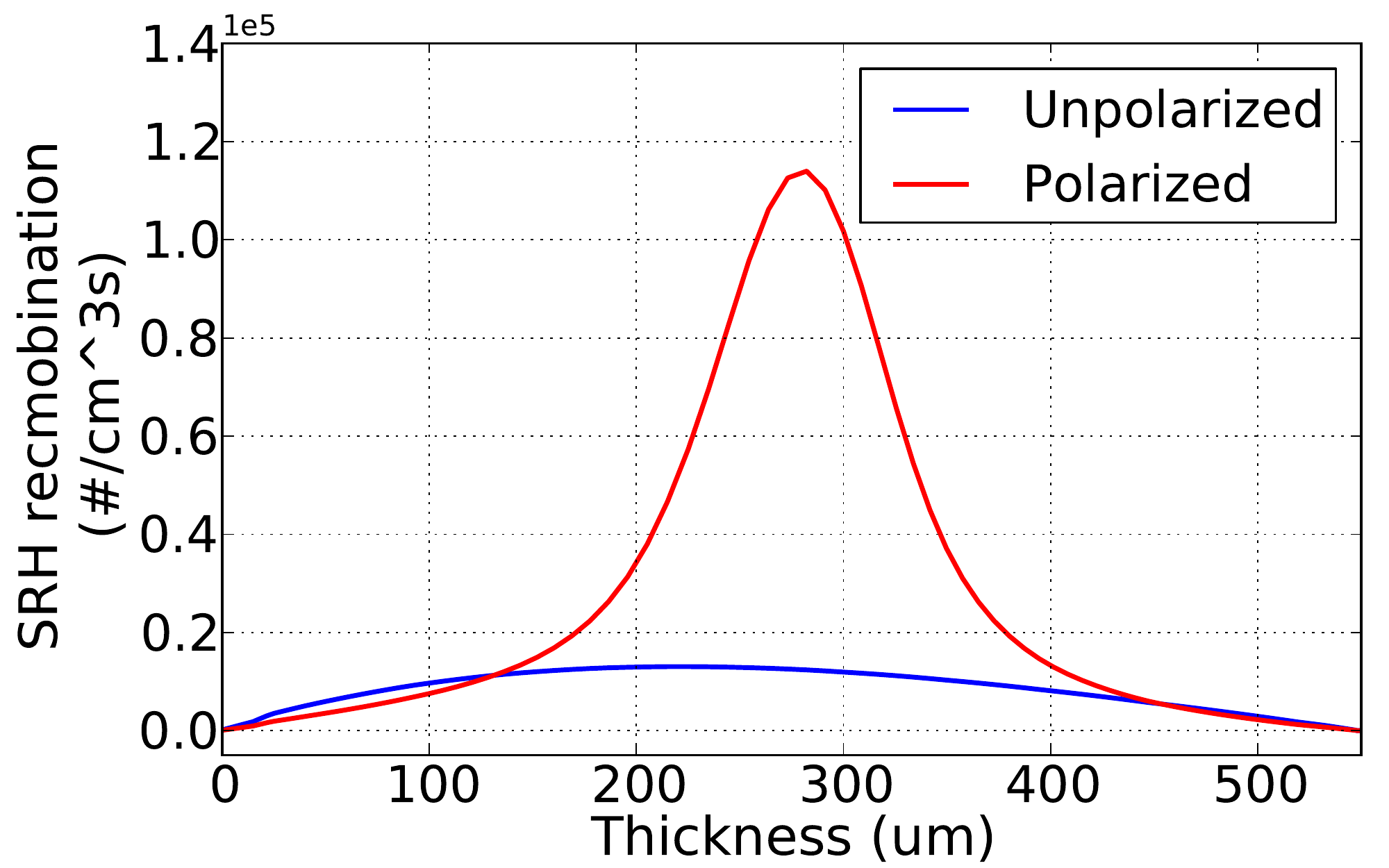}}\hfill
\subfloat[Electrical field at $200\,\rm{V}$.]{%
\includegraphics*[width=.3\textwidth]{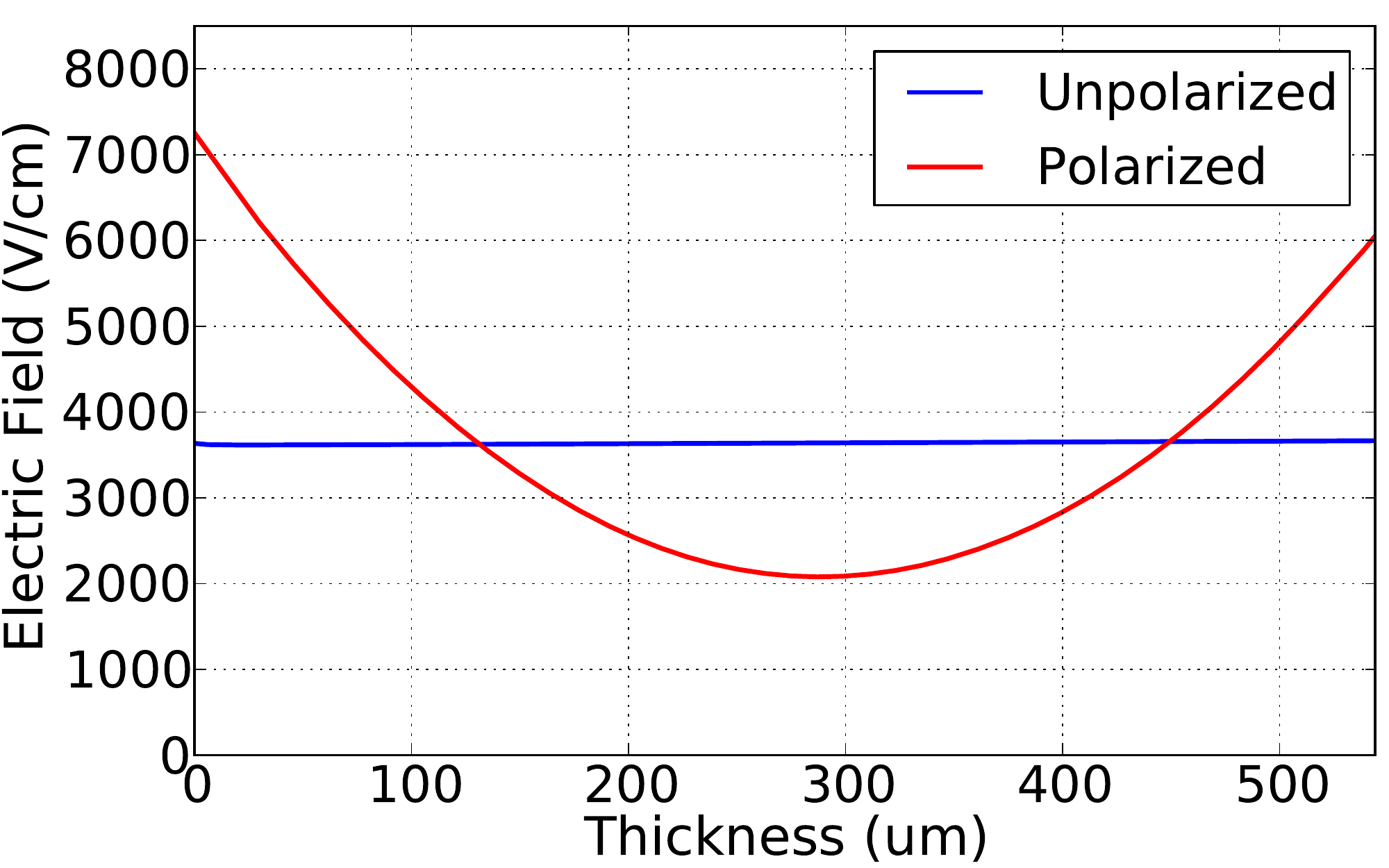}}\hfill
\caption{%
Comparison of the TCT pulse modification for the unpolarized (blue) and the polarized (red) diamond state for a diamond sensor irradiated up to a fluence of $f_2$.}
\label{ElectricalField_100V}
\end{figure*}

\begin{figure*}[bth]%
\subfloat[CCE at bias voltages of $100\,\rm{V}$ and $200\,\rm{V}$.]{%
\includegraphics*[width=.3\textwidth]{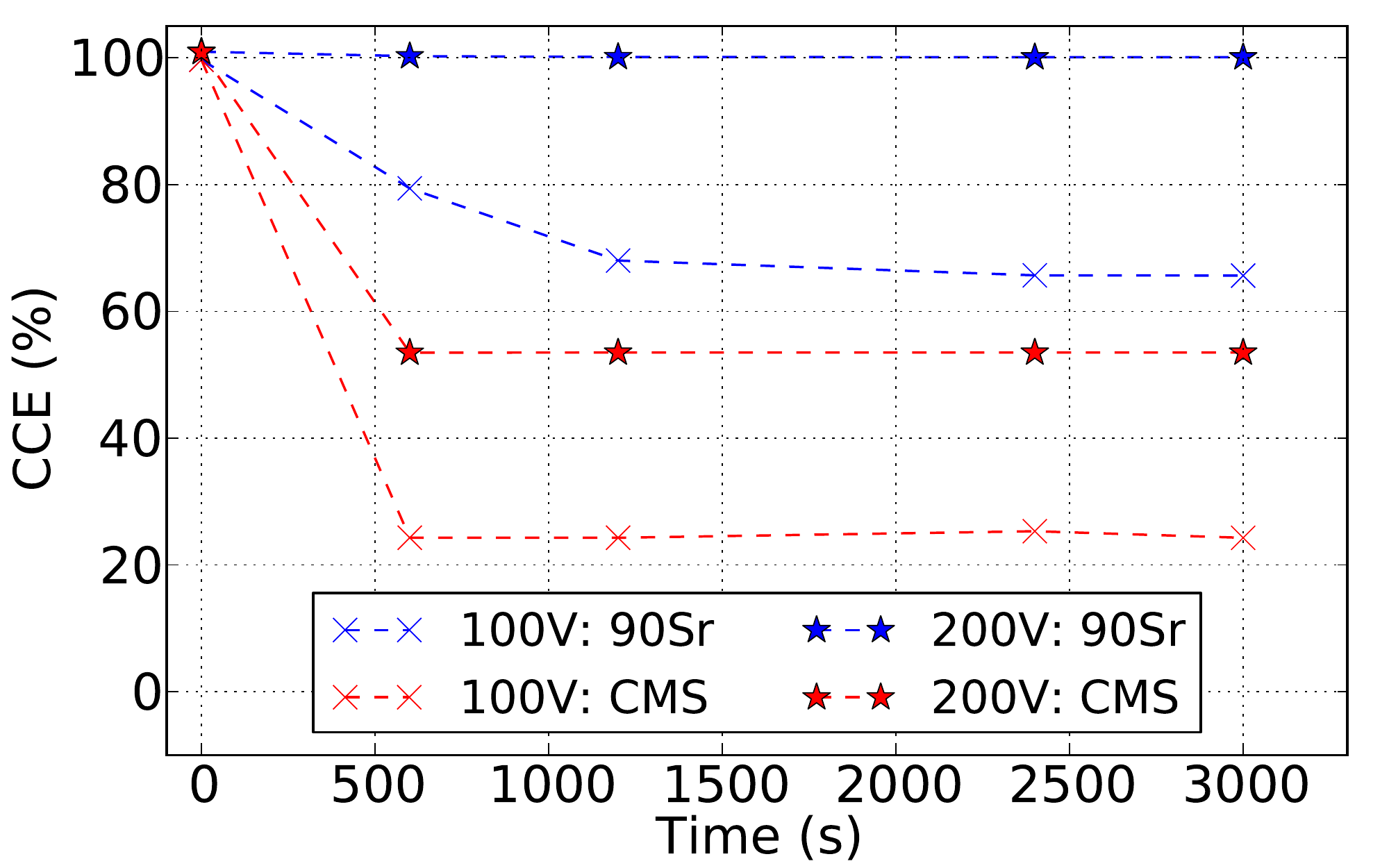}}\hfill
\subfloat[Electrical field at a bias voltage of $100\,\rm{V}$.]{%
\includegraphics*[width=.3\textwidth]{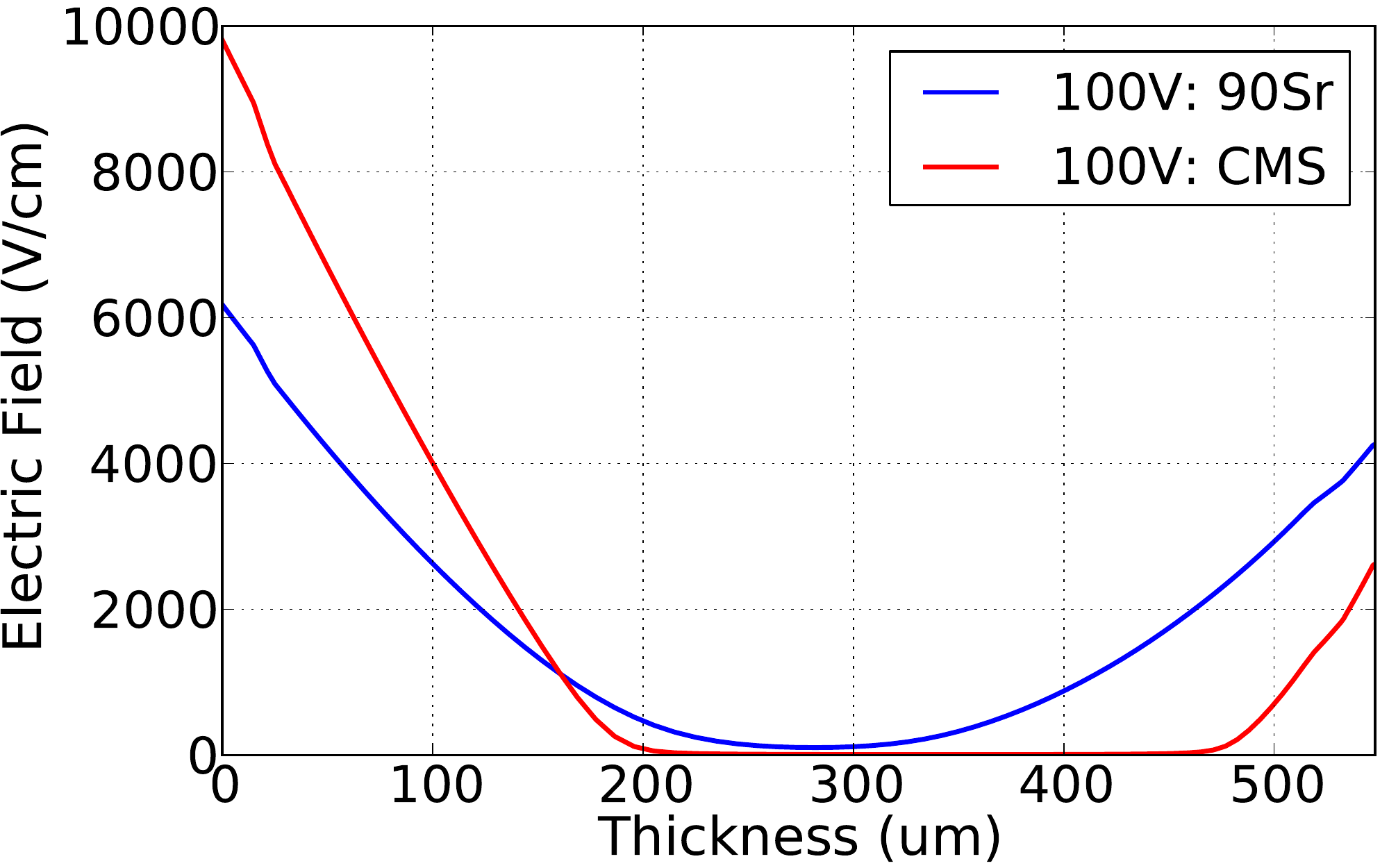}}\hfill
\subfloat[Electrical field at a bias voltage of $200\,\rm{V}$.]{%
\includegraphics*[width=.3\textwidth]{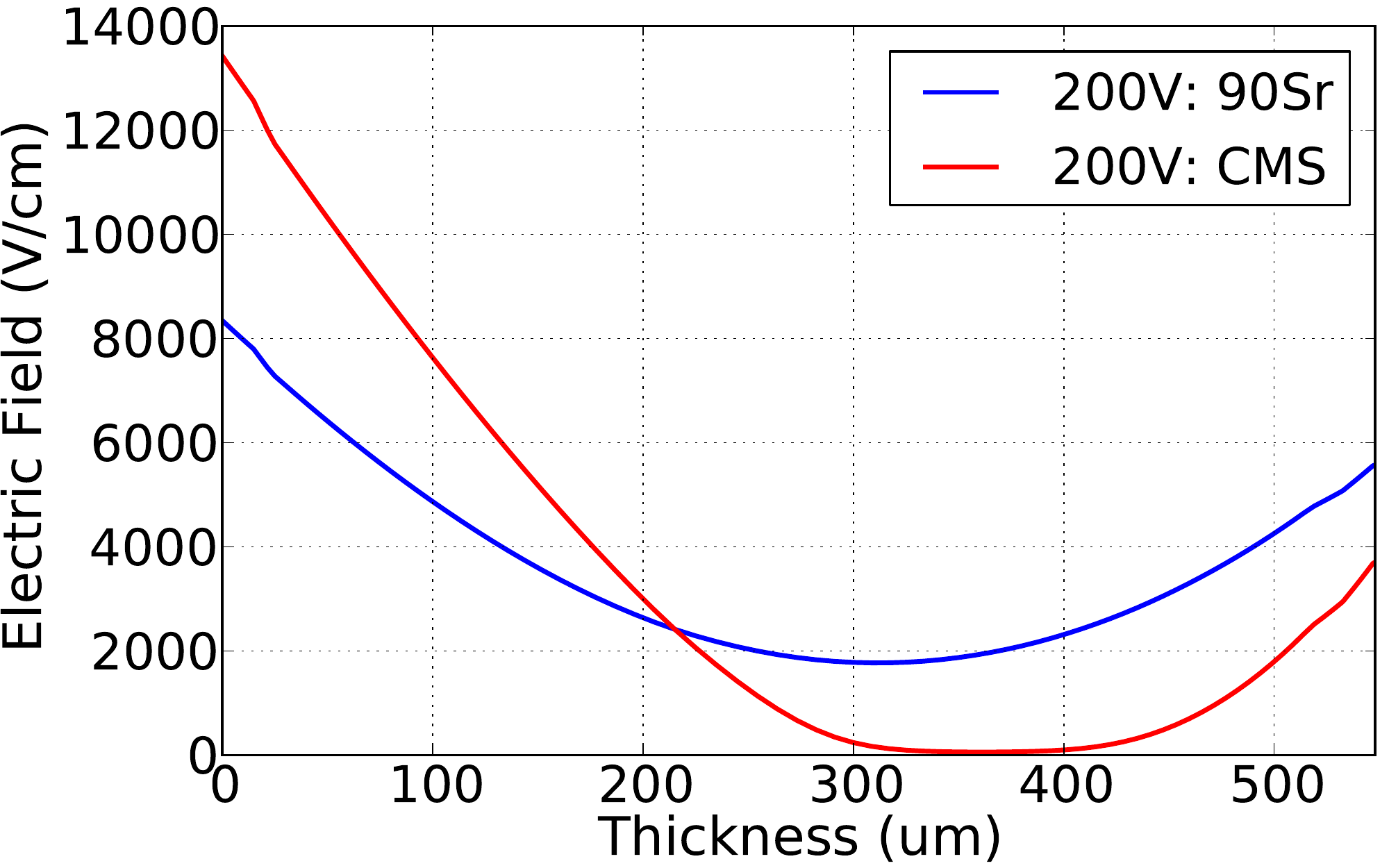}}\hfill
\caption{%
  Comparison of the simulation between the $^{90}Sr$ particle rate (blue) and the CMS particle rate environment (red) for the bias voltages of $100\,\rm{V}$ and $200\,\rm{V}$.}
\label{CCE_CMS}
\end{figure*}

\subsection{Simulation versus experimental data}
The SILVACO TCAD simulation follows the experimental measurement procedure and simulates the same time duration of the experimental measurement, in total $3600\,\rm{s}$. The diamond sample is exposed during the entire simulation to the FLUKA simulated $^{90}Sr$ source.

In a first step the diamond is exposed to the $^{90}Sr$ source over a time duration of 20\,minutes with no bias voltage applied in order to fill the diamond traps homogeneously. In a next simulation step the bias voltage is quickly ramped up ($<1\,\rm{s}$) and followed by $\alpha$ particle hits simulated on both surfaces successively to simulate the TCT pulse of the electron and the hole drift. This is followed by a MIP particle hit used to calculate the CCE. This procedure of three particle hits is simulated at different time steps ($0\,\rm{s}$, $300\,\rm{s}$, $600\,\rm{s}$, $1200\,\rm{s}$, $1800\,\rm{s}$, $2400\,\rm{s}$, $3000\,\rm{s}$ and $3600\,\rm{s}$) within the entire simulation time in order to probe the TCT pulses and the CCE. At these measurement steps additional diamond information like electrical field and recombination rates are probed as well.

The simulation of the TCT pulse modification at 3 different time steps for the electron drift is compared with the experimental data in Fig.~\ref{TCT_100V_ele_full}. Beside the correct description of the TCT pulses the evolution in time is in agreement with the measurement, simulating well the build up of polarization. In Fig.~\ref{Simulation_100V} the simulation result for the hole and electron drift at different bias voltages are compared to the experimental results for the unpolarized and polarized diamond state. The CCE simulation results compared to the experimental data are shown in Figs.~\ref{Simulation_100V}c,f,i. All TCT and CCE simulation results are in agreement with the experimental data concerning the applied bias voltage and the evolution in time.

\paragraph{Analysis of the electrical field}
The electrical fields corresponding to the unpolarized and polarized TCT pulse shapes are shown in Fig.~\ref{ElectricalField_100V}a for a bias voltage of $100\,\rm{V}$. The rectangular TCT shape (Fig.~\ref{Simulation_100V}a,b - unpolarized) corresponds to a constant electrical field ($\overrightarrow{E}=1800\,\rm{V/cm}$). The corresponding electrical field for the polarized diamond state (Fig.~\ref{Simulation_100V}a,b - polarized) however has a local minimum in the middle of the diamond bulk. In the mi\-nimum the electric field ($\overrightarrow{E}=300\,\rm{V/cm}$) is a factor six reduced compared with the unpolarized state. This locally reduced electrical field is increasing the recombination rate of the charge carriers by a factor of 10, see Fig.~\ref{ElectricalField_100V}b, thus explaining the drop from $93\,\%$ to $65\,\%$ in CCE.

The influence of the polarization for an increased bias voltage of $200\,\rm{V}$ is reduced, thus leading to a less affected electrical field, as shown in Fig.~\ref{ElectricalField_100V}c.

\section{Extrapolation of the simulation to the CMS particle environment}
The effective recombination center model was fitted to laboratory particle rates created by a $^{90}Sr$ source. However, in the CMS detector the particle rate is about a factor 30 higher.

At a bias voltage of $100\,\rm{V}$ this high particle rate leads to a stronger reduction in CCE, from 65\,\%~($^{90}Sr$) to 23\,\%~(CMS), as shown in Fig.~\ref{CCE_CMS}a. For an increased bias voltage of $200\,\rm{V}$ the CCE is reduced from 100\,\%~($^{90}Sr$) to 51\,\%~(CMS). This reduced CCE can be explained by the corresponding electrical fields, shown in Figs.~\ref{CCE_CMS}b and \ref{CCE_CMS}c. At a bias voltage of $100\,\rm{V}$ almost $50\,\%$ of the diamond bulk has an electrical field close to zero.

\section{Conclusion}
The charge collection efficiency in the diamond sensors used in the beam loss monitors for the LHC turned out to be far below expectation after a rather low irradiation level. Detailed investigations pointed to a decrease in the electric field inside the sensor by the space charges trapped in the defects from the radiation damage, called polarization. So the defects not only reduce the CCE by trapping the charge, but the reduced electric field enhances the recombination of charges, thus reducing the CCE in a strongly non-linear way. These interpretations were supported by a detailed simulation of the TCT and CCE measurements using an effective recombination center model with the energy levels and trapping cross sections fitted to the data obtained from laboratory measurements with a $^{90}Sr$ source. Extrapolating the reduction of the electric field by the polarizing space charge inside the sensor to the high rate environment of the CMS detector explained the poor performance of the diamond sensors in the harsh environment of the LHC.

A possibility to avoid the diamond polarization in a high particle rate environment could be the switching of the bias voltage with a few Hz, so the space charge would switch direction as well and could not build up so strongly. Alternatively, one should try to increase the high voltage breakthrough voltage, so one could operate with an electric field from the bias voltage well above the electric field from the space charge.

\begin{acknowledgement}
This work has been sponsored by the Wolfgang Gentner Programme of the Federal Ministry of Education and Research and been supported by the H2020 project AIDA-2020, GA no. 654168 (http://aida2020.web.cern.ch/).
\end{acknowledgement}

%

\begin{thebibliography}{[1]}

\bibitem{Guthoff2013168}%
 M.~Guthoff et al., Nuclear Instruments and Methods in Physics Research Section A: Accelerators, Spectrometers, Detectors and Associated Equipment \textbf{730}, 168 - 173 (2013).
 
\bibitem{Guthoff2015}%
 M.~Guthoff, W.~de~Boer, A.~Dabrowski, F.~Kassel and D.~Stickland, PoS: Proceedings - 3rd International Conference on Technology and Instrumentation in Particle Physics (TIPP 2014) \textbf{281}, (2014).
 
\bibitem{Guthoff2014}%
 M.~Guthoff, Radiation damage to the diamond based Beam Condition Monitor of the CMS Detector at the LHC, PhD thesis IEKP-KA/2014-01 (2014). 
 
\bibitem{RD42}%
 RD42~Coll. (W.~Adam et al.), Nuclear Instruments and Methods in Physics Research Section A: Accelerators, Spectrometers, Detectors and Associated Equipment \textbf{565}, 278 - 283 (2006).
 
\bibitem{deBoer2007}%
 W.~de~Boer et al., Physica Status Solidi (a) \textbf{204}, 3004-3010 (2007).
 
\bibitem{Guthoff2013}%
 M.~Guthoff, W.~de~Boer and S.~M\"{u}ller, Nuclear Instruments and Methods in Physics Research A \textbf{735}, 223-228 (2014).

\bibitem{Rebai2016}%
 M.~Rebai et al, Diamond and Related Material \textbf{61}, 1-6 (2016).

\bibitem{Valentin2015}%
 A.~Valentin et al, Physica Status Solidi (a) \textbf{212}, 2636-2640 (2015).

\bibitem{Element6}%
 ElementSix: Synthetic diamond producer, www.e6.com.
 
\bibitem{NeutronFacility}%
 A.~Kolšek, V.~Radulovi\'{c}, A.~Radulovi\'{c} and L.~Snoj, Nuclear Engineering and Design \textbf{283}, 155 - 161 (2015).

\bibitem{Cindro2015}%
 S.~Lagomarsino et al, Applied Physics Letters \textbf{106}, 193509 (2015)

\bibitem{Isberg2002}%
 J.~Isberg et al, Science \textbf{297}, 1670 (2002)

\bibitem{Pernegger2005}%
 H.~Pernegger et al, Journal of Applied Physics \textbf{97}, 073704 (2005)

\bibitem{Ramo1939}%
 S.~Ramo, Proceedings of the IRE \textbf{27 - 9}, 584 - 585 (1339).

\bibitem{Particulars_Amp}%
 Particulars, Advanced Measurement Systems: Particulars Wide Band Current Amplifier, www.particulars.si.

\bibitem{Grah2009}%
 C.~Grah et al., IEEE Transactions on Nuclear Science \textbf{56 - 2}, 462 - 467 (2009).
  
\bibitem{Pomorski}%
 M.~Pomorski, Physica Status Solidi (a) \textbf{212}, 2553-2558 (2015).
 
\bibitem{Caughey67}%
 D.~M.~Caughey and R.~E.~Thomas, Proceedings of the IEEE \textbf{55 - 12}, 2192-2193 (1967).
 
\bibitem{Silvaco}%
 Silvaco TCad - Atlas: Semiconductor device simulator; www.silvaco.com.
 
\bibitem{FLUKA1}%
 A.~Ferrari, P.~R.~Sala, A.~Fasso and J.~Ranft,'FLUKA: a multi-particle transport code', \textbf{CERN 2005-10}, SLAC-R-773 (2005).
 
\bibitem{FLUKA2}%
 T.~T.~Boehlen et al, 'The FLUKA Code: Developments and Challenges for High Energy and Medical Applications', Nuclear Data Sheets \textbf{120}, 211-214 (2014)
 
\bibitem{SRH1952}%
 W.~Shockley and W.~T.~Read, Physical Review \textbf{87}, 835-842 (1952).
 
\bibitem{ATALASmanual}%
 Silvaco, Atlas User Manual - Device simulation software (2010).

\end{thebibliography}
%

\end{document}